\documentclass[12pt]{article}

\usepackage{fullpage}
\usepackage{authblk}
\usepackage[american]{babel}
\usepackage[utf8]{inputenc}		
\usepackage[T1]{fontenc}	

\usepackage[round]{natbib}
\usepackage{enumerate}
\usepackage{graphicx}		
\usepackage{epstopdf}
\usepackage{marvosym}		
\usepackage{amsmath}			
\usepackage{amssymb}
\usepackage{amsthm}
\usepackage{amsfonts}
\usepackage{mathtools}
\usepackage{xcolor}
\usepackage{float}
\usepackage{url}
\usepackage[format=plain,labelfont=bf]{caption}
\usepackage{subcaption}
\usepackage{sidecap}
\usepackage{capt-of}
\usepackage{booktabs}
\newcommand{\bigcell}[2]{\begin{tabular}{@{}#1@{}}#2\end{tabular}}
\usepackage{longtable}
\usepackage{multirow}
\usepackage{rotating}
\usepackage{wrapfig}
\usepackage[english=american]{csquotes}
\usepackage[hidelinks]{hyperref}
\usepackage[parfill]{parskip}
\usepackage[normalem]{ulem}
\usepackage{textcomp}

\usepackage{imakeidx}
\makeindex

\newcommand\blfootnote[1]{%
  \begingroup
  \renewcommand\thefootnote{}\footnote{#1}%
  \addtocounter{footnote}{-1}%
  \endgroup
}

\theoremstyle{definition}
\newtheorem{definition}{Definition}

\theoremstyle{plain}

\newcommand{\cN}{{\mathcal N}}
\newcommand{\cT}{{\mathcal T}}

\newcommand{\splitatcommas}[1]{%
  \begingroup
  \ifnum\mathcode`,="8000
  \else
    \begingroup\lccode`~=`, \lowercase{\endgroup
      \edef~{\mathchar\the\mathcode`, \penalty0 \noexpand\hspace{0pt plus 1em}}%
    }\mathcode`,="8000
  \fi
  #1%
  \endgroup
}

\newcommand{\tuple}[1]{(\splitatcommas{#1})}

\title{Classes of Explicit Phylogenetic Networks and their Biological and Mathematical Significance}
\author[1,$\dagger$]{Sungsik Kong}
\author[2,$\dagger$]{Joan Carles Pons}
\author[1,3]{Laura Kubatko}
\author[4,$\ast$]{Kristina Wicke}

\affil[1]{Department of Evolution, Ecology, and Organismal Biology, The Ohio State University, Columbus OH, USA}
\affil[2]{Department of Mathematics and Computer Science, University of the Balearic Islands, Palma 07122, Spain}
\affil[3]{Department of Statistics, The Ohio State University, Columbus OH, USA}
\affil[4]{Department of Mathematics, The Ohio State University, Columbus OH, USA}

\date{}

\begin{document}

\maketitle

\begin{abstract}
\noindent The evolutionary relationships among organisms have  traditionally been represented using rooted phylogenetic trees. However, due to reticulate processes such as hybridization or lateral gene transfer, evolution cannot always be adequately represented by a phylogenetic tree, and rooted phylogenetic networks that describe such complex processes have been introduced as a generalization of rooted phylogenetic trees. In fact, estimating rooted phylogenetic networks from genomic sequence data and analyzing their structural properties is one of the most important tasks in contemporary phylogenetics. Over the last two decades, several subclasses of rooted phylogenetic networks (characterized by certain structural constraints) have been introduced in the literature, either to model specific biological phenomena or to enable tractable mathematical and computational analyses. In the present manuscript, we provide a thorough review of these network classes, as well as provide a biological interpretation of the structural constraints underlying these networks where possible. In addition, we discuss how imposing structural constraints on the network topology can be used to address the scalability and identifiability challenges faced in the estimation of phylogenetic networks from empirical data.
\end{abstract}

\textit{Keywords:} hybridization, introgression, lateral gene transfer, phylogenetic tree, phylogenetic network

\blfootnote{$^\dagger$These authors contributed equally to this work. \\ $^\ast$Corresponding author\\ \textit{Email address:} \url{wicke.6@osu.edu}}

\section{Introduction}
Reconstructing and analyzing the evolutionary relationships among organisms is a central goal in evolutionary biology.  Rooted phylogenetic trees (often simply referred to as \emph{phylogenies}) have traditionally been used to represent the evolutionary history for a collection of taxa. In a rooted phylogenetic tree, the leaves (or terminal vertices) usually represent sampled extant taxa, the root corresponds to the most recent common ancestor of the taxa under consideration, and all other interior vertices can be interpreted as split (or speciation) events, where some ancestral taxon evolved into two (or more) distinct taxa. In particular, rooted phylogenetic trees assume vertical inheritance, where genomic material is transmitted from an ancestral species to a descendant species.

\begin{figure}[htbp]
    \centering
    \begin{subfigure}[b]{0.45\textwidth}
         \centering
         \includegraphics[width=\textwidth]{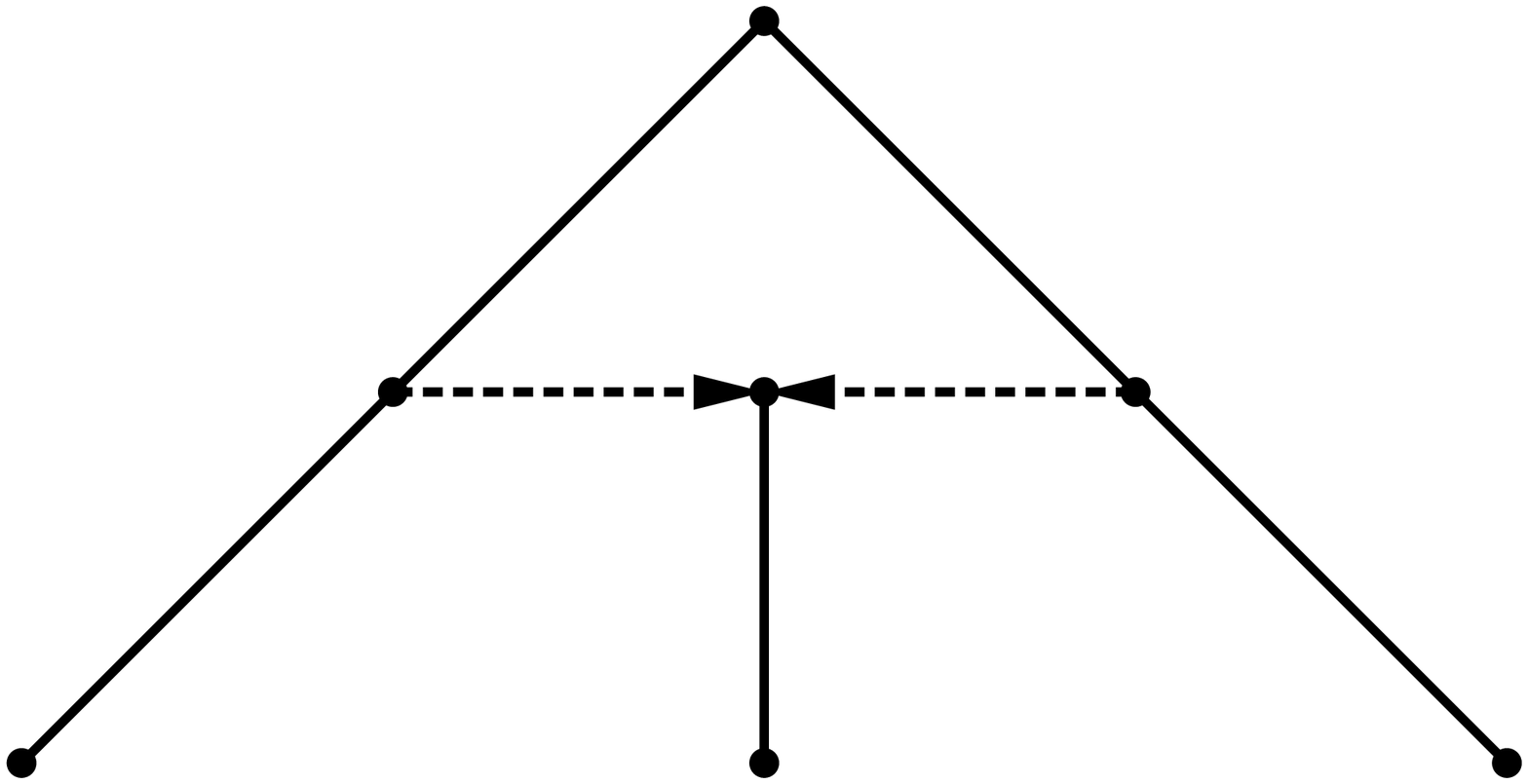}
         \caption{Hybrid speciation}
     \end{subfigure}
     \begin{subfigure}[b]{0.45\textwidth}
         \centering
         \includegraphics[width=\textwidth]{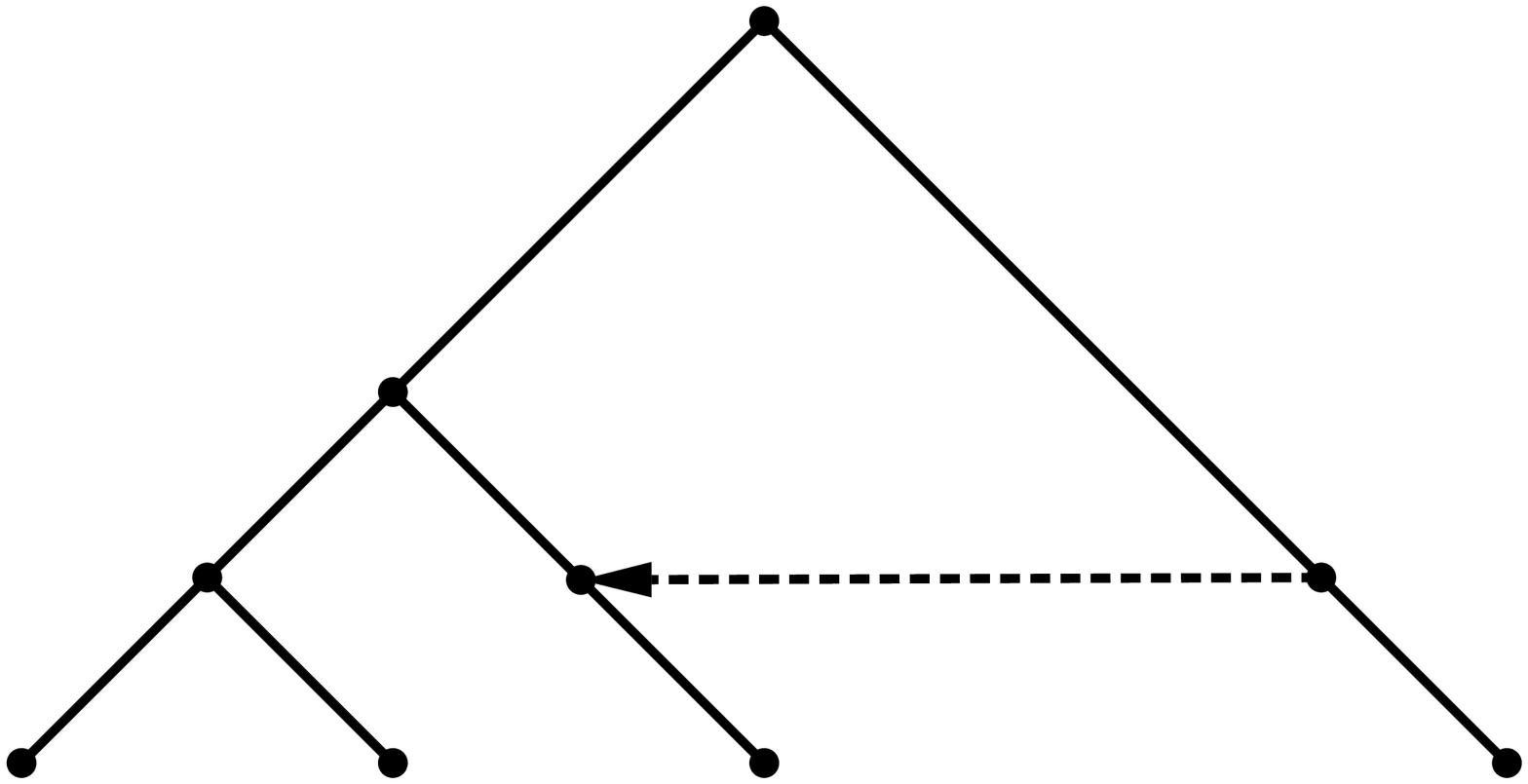}
         \caption{Introgression/LGT}
     \end{subfigure}
    \caption{(a) Schematic representation of hybrid speciation, where the horizontal dotted lines indicate interbreeding from two distinct ancestral populations that leads to formation of a hybrid taxon. (b) Schematic representation of introgression/LGT, where the right-most lineage transfers genetic information to another lineage at the point in the evolutionary history indicated by the directed horizontal dotted edge.}
    \label{Fig_Reticulation}
\end{figure}

However, it is nowadays widely accepted that the evolutionary pathway of an organism is not always tree-like and that many systems in nature experience events in which genetic information is transferred `horizontally' between taxa rather than `vertically'. These events include hybridization\index{hybridization} (which in turn includes hybrid speciation and introgression~\citep{anderson1953}), lateral gene transfer (LGT; sometimes also called horizontal gene transfer (HGT)), and recombination.
Briefly, hybrid speciation\index{hybridization!hybrid speciation} refers to the emergence of a novel lineage through interbreeding between two distinct parental lineages. Introgression\index{hybridization!introgression} describes the transmission of genetic information from one lineage (the donor) to another lineage (the recipient) by repeated backcrossing of a hybrid daughter with one of its parents, whereas LGT\index{lateral gene transfer} usually refers to the transmission of genetic material from a donor to a recipient via processes such as transformation, transduction, or conjugation (see Figure~\ref{Fig_Reticulation} for a schematic representation of hybrid speciation and introgression/LGT). Finally, recombination\index{recombination} refers to a rearrangement of genetic material through crossing over of chromosomes during reproduction between a pair of individuals in the same lineage. Note that recombination is thus an intraspecific process, whereas hybrid speciation and LGT are interspecific processes.

Because phylogenetic trees are not adequate to represent non-treelike evolutionary histories such as those described above, rooted phylogenetic networks have been proposed as a generalization of rooted phylogenetic trees in the literature. Rooted phylogenetic networks are often referred to as explicit (directed) networks since they explicitly depict the evolution of a group of organisms from a common ancestor via a combination of splitting and reticulation events. In brief, an explicit phylogenetic network is a rooted directed graph with all edges directed away from the root (see Definition~\ref{Def_PhylogeneticNetwork} for a formal definition). It is important to distinguish explicit networks from implicit or abstract networks such as split networks~\citep{Bandelt1992} or median-joining networks~\citep{Bandelt1999} that depict phenetic relatedness based on overall similarity among taxa but do not represent information on the evolutionary history or direction of evolution~\citep{kong2016,sanchez-pacheco2020}. 
More recently, an `intermediate' class of phylogenetic networks, semi-directed phylogenetic networks, was introduced in the literature (e.g.,~\cite{solis-lemus2016}). Roughly, semi-directed phylogenetic networks are obtained from rooted phylogenetic networks by suppressing the root and ignoring the direction of all edges, except for those involved in a reticulation event, thus keeping information on which vertices are reticulations.

In this paper, we focus on explicit networks as these networks directly model the evolutionary history of a collection of species, rather than simply displaying species similarity, and thus have direct biological relevance. 
The reconstruction and analysis of these networks is one of the most active areas of research in computational or mathematical phylogenetics and a variety of subclasses of rooted phylogenetic networks have been introduced in the literature. Some of these classes were introduced to explicitly model specific evolutionary scenarios, whereas others seem to have been mainly established for mathematical convenience or computational tractability.

We aim at providing a thorough review of these network classes, as well as an analysis of which are biologically interpretable (in the sense that they depict realistic evolutionary scenarios and are expected to capture features of the true evolutionary history for empirical data sets) and which are merely of mathematical interest. In addition, we discuss how imposing structural constraints on the networks can address mathematical and computational challenges faced when estimating phylogenetic networks from data. Here, we focus in particular on the notions of scalability and identifiability. By {\itshape scalability}, we mean how the performance of the inference method, in terms of both computational time and accuracy, changes as the size of the problem (e.g., the number of taxa, the amount of available data, or both; or the complexity of the network, e.g., the number of reticulations in the network) increases. By {\itshape identifiability}, we are referring to the study of which features of the phylogenetic network are uniquely determined by the model from which the data arise or are uniquely characterized by certain substructures of the network like `displayed' trees. 

To our knowledge, this study is the first comprehensive review of the more than 20 classes of explicit phylogenetic networks discussed in the literature that not only summarizes their structural properties but also establishes a connection to biological processes (where possible), discusses some scalability challenges faced in network estimation in practice and reviews identifiability results obtained for different classes. We remark, however, that there are some excellent books or book chapters on the general topic of phylogenetic networks and related concepts~\citep[e.g., ][]{Huson2011,Morrison2011,Gusfield2014,Steel2016,Zhang2019a} as well as some web resources such as the websites \enquote{Who is Who in Phylogenetic Networks}\footnote{\url{https://phylnet.univ-mlv.fr/}} \citep{Agarwal2016} and \enquote{ISIPhyNC (Information System on Inclusions of Phylogenetic Network Classes)}\footnote{\url{https://phylnet.univ-mlv.fr/isiphync/index.php}} \citep{Gambette2018a}.

Note, however, that we restrict our review to binary phylogenetic networks, in which each speciation event leads to precisely two new species and each reticulation event involves exactly two parental species. On one hand, rooted binary phylogenetic networks are the most common type of explicit phylogenetic networks studied in the literature. On the other hand, we are particularly interested in the biological meaning of the networks under consideration and it is biologically unlikely that a speciation event results in three or more new lineages or that a reticulation event involves three or more distinct species. Nevertheless, we remark that rooted non-binary (or multifurcating) phylogenetic networks can, for example, be useful to reflect uncertainty in the order of speciation or reticulation events. 

The remainder of this paper is organized as follows. We begin by introducing and discussing the most important concept of this paper, namely rooted binary phylogenetic networks, in Section~\ref{Sec_NetworkDef}. In addition, we introduce additional terminology and notation used throughout this manuscript. In Section~\ref{Sec_NetworkClasses}, we then provide a comprehensive review of more than 20 classes of rooted binary phylogenetic networks currently used in the literature. We first describe classes of networks with an underlying biological interpretation (Section \ref{Subsec_meaningful}), before discussing additional classes not linked to biology yet (Section \ref{Subsec_notlinked}).
Afterwards, in Section~\ref{Sec_ScalabilityIdentifiability}, we discuss two important mathematical and computational challenges in estimating phylogenetic networks from data. More precisely, we consider the notions of scalability and identifiability. In particular, we describe why and how scalability and identifiability issues affect phylogenetic network estimation in practice. We also review approaches used to address these challenges, and we summarize positive and negative results obtained, highlighting the impact of structural constraints on the estimation of networks. We end with a brief discussion in Section~\ref{Sec_Conclusion}.

\section{Rooted binary phylogenetic networks and related concepts} \label{Sec_NetworkDef}
Before we can review and discuss various classes of rooted binary phylogenetic networks and discuss their biological relevance, we need to provide a formal definition and to introduce some additional definitions and concepts.

\paragraph{Phylogenetic networks and phylogenetic trees.}
Throughout the paper, $X$ denotes a non-empty finite set (of taxa). 

\begin{definition}[Rooted binary phylogenetic network]\label{Def_PhylogeneticNetwork} \index{phylogenetic network}\index{phylogenetic network!binary}
A \emph{rooted binary phylogenetic network $\cN=(V,E)$ on $X$ with root $\rho$} is a rooted directed acyclic graph with no parallel arcs satisfying the following properties:
\begin{enumerate}[(i)]
    \item the (unique) root $\rho$ has in-degree zero and out-degree two;
    \item a vertex with out-degree zero has in-degree one, and the set of vertices with out-degree zero is identified with $X$;
    \item all other vertices have either in-degree one and out-degree two, or in-degree two and out-degree one.
\end{enumerate}
\end{definition}

\begin{figure}[htbp]
    \centering
    \includegraphics[scale=0.25]{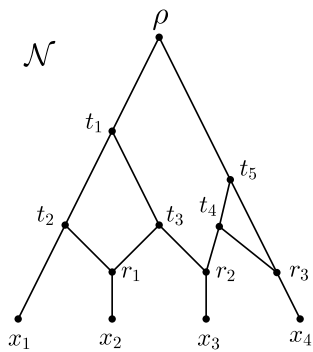}
    \caption{Rooted binary phylogenetic network $\cN$ on $X=\{x_1, \ldots, x_4\}$. All arcs are directed down the page beginning from the root $\rho$. Vertices labelled $t_{\ast}$ are tree vertices of $\cN$ and vertices labelled $r_{\ast}$ are reticulation vertices. Arcs $(t_{\ast}, r_{\ast})$ are reticulation arcs and all others are tree arcs. 
    Apart from $t_3$ and $t_4$ all vertices of $\cN$ are visible. Moreover, the arc $(t_5,r_3)$ is a shortcut as there is also a directed path from $t_5$ to $r_3$ in $\cN$ (i.e., the path $(t_5, t_4,r_3)$).}
    \label{Fig_NetworkConcepts}
\end{figure}

For technical reasons, if $\vert X \vert=1$, we allow $\cN$ to consist of the single vertex in $X$. Moreover, the vertices in $X$ are called \emph{leaves}\index{leaf}, the vertices with in-degree one and out-degree two are called \emph{tree vertices}\index{tree vertex}, and the vertices with in-degree two and out-degree one are called \emph{reticulation vertices}\index{reticulation vertex}. All arcs directed into a reticulation vertex are called \emph{reticulation arcs}\index{reticulation arc} and all other arcs are called \emph{tree arcs}\index{tree arc}. An example of a rooted binary phylogenetic network is depicted in Figure~\ref{Fig_NetworkConcepts}. Here, as in all subsequent figures in the paper, arcs are directed down the page. 

Two networks $\cN_1=(V_1,E_1)$ and $\cN_2=(V_2,E_2)$ on $X$ are said to be \emph{isomorphic} if there exists a bijection $\varphi: V_1 \rightarrow V_2$ such that $\varphi(x)=x$ for all $x \in X$, and $(u,v)$ is an arc in $\cN_1$ if and only if $(\varphi(u), \varphi(v))$ is an arc in $\cN_2$.

A \emph{rooted binary phylogenetic $X$-tree $\cT$}\index{phylogenetic tree} is a rooted binary phylogenetic network on $X$ with no reticulation. Unless explicitly stated otherwise, when we refer to phylogenetic networks (trees), we will always mean rooted binary phylogenetic networks (trees).

Finally, note that for most parts of the paper we are interested in structural properties of different phylogenetic networks, i.e., we are interested in network \emph{topologies} or \emph{shapes}. However, when discussing the concepts of scalability and identifiability in Section~\ref{Sec_ScalabilityIdentifiability}, we will also mention branch lengths\index{branch lengths} and inheritance probabilities\index{inheritance probability}. In this case, we assume that each arc of $\cN$ has a non-negative real-valued length, i.e., we assume that there exists a mapping $w: E \rightarrow \mathbb{R}_{\geq 0}$ under which each arc $e$ of $\cN$ is assigned  the weight $w(e)$. Additionally, if $v$ is a reticulation vertex with in-coming reticulation arcs $e_1=(u_1,v)$ and $e_2=(u_2,v)$, we assume that $e_1$ and $e_2$ are associated with probabilities $\gamma_{e_1} \in (0,1)$ and $\gamma_{e_2} = 1- \gamma_{e_1}$ that indicate the proportional contribution of genetic material or features that vertex $v$ inherits from its parents $u_1$ and $u_2$, respectively.\label{inheritanceprob}

\paragraph{Inherent assumptions underlying Definition~\ref{Def_PhylogeneticNetwork}.}
Before discussing additional concepts related to phylogenetic networks, we want to point out some inherent assumptions about the evolutionary history of a group of organisms that underlie Definition~\ref{Def_PhylogeneticNetwork}. First, as indicated in the introduction, phylogenetic networks in the sense of this definition depict evolution as a directed process starting at the root of the network and moving towards its leaves. Here, the root corresponds to the most recent common ancestor of the organisms under consideration, and this ancestral organism is assumed to be unique. However, the phylogenetic network does not contain any information about the ancestry of the root itself. Moreover, it is assumed that all taxa under consideration correspond to the leaves of the network, whereas all internal vertices represent hypothetical ancestral species. In particular, it is assumed that all data observed today is observed at the leaves of the network. 

Additionally, Definition~\ref{Def_PhylogeneticNetwork} establishes a bijection between the taxa under consideration (represented by the set $X$) and the set of leaves of the network (since the two sets are identified). This means that a particular leaf of a phylogenetic network represents precisely one species and a particular species is represented by precisely one leaf. In other words, multi-labelling of leaves (i.e., two or more leaves representing the same species) is excluded. However, there is a close relationship between phylogenetic networks and multi-labelled trees (MUL-trees)\index{MUL-tree}, leaf-labelled trees where more than one leaf may have the same label, that for example arise in the study of polyploids (organisms having multiple complete copies of their genome). In particular, it has been shown that a phylogenetic network can be `unfolded' to obtain a MUL-tree, and a MUL-tree can under certain conditions be `folded' into a phylogenetic network that exhibits it \citep{Huber2006, Huber2016a,Huber2020}. 

Third, and again as indicated in the introduction, Definition~\ref{Def_PhylogeneticNetwork} imposes certain constraints on the degrees of vertices of the network. Tree vertices\index{tree vertex} have an out-degree of precisely two, implying that each speciation events leads to precisely two new species. In other words, \emph{polytomies}, i.e., vertices with an out-degree of three or greater which can either be interpreted as uncertainty in the order of speciation events (soft polytomies) or simultaneous divergence of three or more species (hard polytomies) are excluded from the definition. Similarly, reticulation vertices\index{reticulation vertex} have an in-degree of exactly two, implying that exactly two parental species are involved in a reticulation event. In addition, all reticulation vertices have an out-degree of one. This can be seen as a way of representing the reticulation \emph{process} with the single child of a reticulation vertex representing the \emph{resulting} taxon. Note that by collapsing both vertices (i.e., the reticulation vertex and its child) into a single vertex representing both the reticulation process and the resulting organism, equivalent mathematical models are obtained. We remark, however, that requiring reticulation vertices to have an out-degree of one does not exclude the single child of the reticulation vertex from being a reticulation vertex itself. As an example, the single child of the reticulation vertex $u$ of the phylogenetic network $\cN_2$ depicted in Figure~\ref{Fig_TreeChild} is itself a reticulation vertex. Last, there are no vertices of in-degree less or equal to one and out-degree one (often referred to as \emph{elementary} vertices)\index{elementary vertex}. An elementary vertex would represent a species that has only one descendant, and it is impossible to distinguish this ancestral species from its unique descendant through biological information only.

As a final observation note that both arcs directed into a reticulation vertex are commonly referred to as `reticulation arcs'\index{reticulation arc}. In particular, they are treated symmetrically and are not distinguished from each other. While this assumption is suitable for modelling hybridization\index{hybridization} or recombination\index{recombination} events where both parental organisms play a symmetrical role in producing the resulting organism, it is less suitable when modelling events of LGT\index{lateral gene transfer} or introgression\index{hybridization!introgression}. In this case, it is often assumed that every reticulation vertex has a single (incoming) reticulation arc, whereas the other in-coming arc is a non-reticulation arc. Under these assumptions, a phylogenetic network can be seen as a `backbone tree' composed by non-reticulation arcs representing the main line of evolution with additional arcs, that is reticulation arcs, added to it (\cite{Cardona2015,Francis2015}). We will elaborate on this idea when introducing LGT networks below.

However, we first need to review some additional concepts related to phylogenetic networks that will be of relevance throughout this manuscript.
\paragraph{(Lowest stable) ancestors, descendants, and siblings.}
Let $\cN=(V,E)$ be a phylogenetic network. If there is an arc $e=(u,v)$ in $\cN$, we say that $u$ is a \emph{parent}\index{ancestor!parent} of $v$, and $v$ is \emph{child}\index{descendant!child} of $u$. Note that $u$ is sometimes also called the \emph{tail} of $e$ and $v$ is called the \emph{head} of $e$. More generally, if there is a directed path between $u$ and $v$, we say that $u$ is an \emph{ancestor}\index{ancestor} of $v$, and $v$ is a \emph{descendant}\index{descendant} of $u$. For vertices $u,v \in V$, we write $u \prec_{\cN} v$ if there is a directed path from $u$ to $v$ (and $u \neq v$). In addition, we write $u \preceq_{\cN} v$ if $u = v$ or $u \prec_{\cN} v$. Given a subset $U \subseteq V$ of the vertices of $\cN$, we say that $u \in U$ is a \emph{lowest} vertex in $U$ if there is no $v \in U$ with $u \prec v$. Now, let $U \subseteq V$ be a subset of the vertices of $\cN$. Then a \emph{stable ancestor}\index{ancestor!stable} of $U$ in $\cN$ is a vertex $v \in V \setminus U$ such that every path from the root to a vertex in $U$ contains $v$. Moreover, the (unique) \emph{lowest stable ancestor}\label{LSA}\index{ancestor!lowest stable} of $U$ in $\cN$ is the lowest such vertex and is denoted by $\text{LSA}_{\cN}(U)$. As an example, consider the rooted binary phylogenetic network depicted in Figure~\ref{Fig_NetworkConcepts} and let $U = \{x_1,x_2\}$. Then, $\text{LSA}_{\cN}(U)=t_1$.
Finally, two vertices $u$ and $v$ of $\cN$ are called \emph{siblings}\index{sibling} if they have a common parent.

\paragraph{Visible vertices, clusters, and shortcuts.}
A vertex $v$ of $\cN$ is said to be \emph{visible}\index{visible vertex} if there is a leaf $x \in X$ such that every directed path from the root of $\cN$ to $x$ traverses $v$ (if $v$ is a leaf simply take $x=v$). We also say that $x$ \emph{verifies the visibility of $v$} or \emph{$x$ verifies $v$} for short \citep{bordewich2016}. As an example, all vertices of the network $\cN$ depicted in Figure~\ref{Fig_NetworkConcepts} apart from $t_3$ and $t_4$ are visible. For example, the reticulation vertex $r_1$ is visible because every directed path in $\cN$ from the root to leaf $x_2$ traverses $r_1$. Thus, $x_2$ verifies the visibility of $r_1$.

Given a phylogenetic network $\cN=(V,E)$ on $X$, the \emph{cluster}\index{cluster} associated with a vertex $v \in V$ is the set 
$$ c_{\cN}(v) = \{x \in X: v \preceq_{\cN} x\},$$
that is the subset of leaves that can be reached from $v$. As an example, consider vertex $t_1$ of the network $\cN$ depicted in Figure~\ref{Fig_NetworkConcepts}. Here, $c_{\cN}(t_1) = \{x_1,x_2,x_3\}$.

Lastly, an arc $e=(u,v)$ of a phylogenetic network $\cN$ is called \emph{redundant} or a \emph{shortcut}\index{reticulation arc!shortcut} if there is a directed path in $\cN$ from $u$ to $v$ that does not use $e$. As an example, arc $(t_5,r_3)$ of the phylogenetic network $\cN$ depicted in Figure~\ref{Fig_NetworkConcepts} is a shortcut as there is also a directed path in $\cN$ from $t_5$ to $r_3$ via $t_4$. Note that the shortcut in $\cN$ induces an (undirected) cycle of length three, consisting of the vertices $t_5$, $t_4$, and $r_3$. However, shortcuts can also be part of larger cycles. For instance, the network $\cN_1$ depicted in Figure~\ref{Fig_NearlyConcepts} contains the shortcut $(u,r_4)$, which is part of an (undirected) cycle of length four. In fact, a shortcut can be part of a cycle of arbitrary length.

\paragraph{Edge subdivision and vertex suppression.}
Finally, we need to introduce two operations on graphs that will be relevant for various network classes discussed below, namely the concepts of \emph{subdividing an edge} and \emph{suppressing a vertex}. Let $\cN$ be a phylogenetic network and let $e=(u,v)$ be an arc of $\cN$. Then, we say that we \emph{subdivide}\index{edge subdivision} $e$, by deleting $e$, introducing a new vertex $w$, and adding the arcs $(u,w)$ and $(w,v)$. Note that the new (elementary) vertex $w$ of in-degree one and out-degree one is often also referred to as an \emph{attachment point}\index{elementary vertex}\index{attachment point}. 
Conversely, given a vertex $w$ of in-degree one and out-degree one with adjacent vertices $u$ and $v$, by \emph{suppressing}\index{vertex suppression} $w$ we mean deleting $w$ and its two incident arcs $(u,w)$ and $(w,v)$ and introducing a new arc $(u,v)$.

\section{Classes of rooted binary phylogenetic networks}
\label{Sec_NetworkClasses}

While Definition~\ref{Def_PhylogeneticNetwork} is the most general definition of rooted binary phylogenetic networks, numerous additional topological restrictions and subclasses have been defined in the literature in recent years. Often these restrictions were introduced to make certain computational and mathematical problems more tractable, but not necessarily to model particular evolutionary scenarios. Therefore, these subclasses must be distinguished from subclasses of phylogenetic networks often defined in biological studies where the categorization is based on the biological phenomenon (e.g., hybridization networks, ancestral recombination graphs) or on the algorithms and software used to infer the networks (e.g., methods discussed in Section \ref{Sec_Scalability}). 

\subsection{Classes of rooted binary phylogenetic networks with a link to biological processes} \label{Subsec_meaningful}
We begin by introducing various classes of rooted binary phylogenetic networks that have been linked to certain evolutionary processes in the literature and are thought to be biologically meaningful.

\paragraph{Temporal or time-consistent networks.}\index{phylogenetic network!temporal}\index{phylogenetic network!time-consistent}
A phylogenetic network $\cN=(V,E)$ is called \emph{temporal} or \emph{time-consistent} \citep{Baroni2006}\label{timeconsistent} if there is a function $t: V \rightarrow \mathbb{N}_{\geq 0}$ such that, for each arc $(u,v)$ of $\cN$, the following two properties hold:
\begin{enumerate}[(T1)]
    \item $t(u) = t(v)$ if $(u,v)$ is a reticulation arc;
    \item $t(u) < t(v)$ if $(u,v)$ is a tree arc.
\end{enumerate}

The map $t$ is called a \emph{temporal labeling} of $\cN$. Note that even if $\cN$ is temporal, the temporal labeling is not unique. More precisely, if $t$ is a temporal labeling of $\cN$, then for each $i \in \mathbb{N}_{\geq 0}$, the map $t': V  \rightarrow \mathbb{N}_{\geq 0}$  defined as $t'(u) = t(u)+i$ for all $u \in V$ also satisfies conditions (T1) and (T2). In particular, while time consistency indicates a potential historical scenario of evolution, it does not represent a unique possible time assignment. 

The biological motivation behind this concept, however, is that for a hybridization event to have occurred, the two species involved (along with the hybrid they formed) must have been extant at the same time (T1). Vertical descent from an ancestral species to a descendant species, on the other hand, implies a passage of time (T2).

\begin{figure}[htbp]
    \centering
    \includegraphics[scale=0.25]{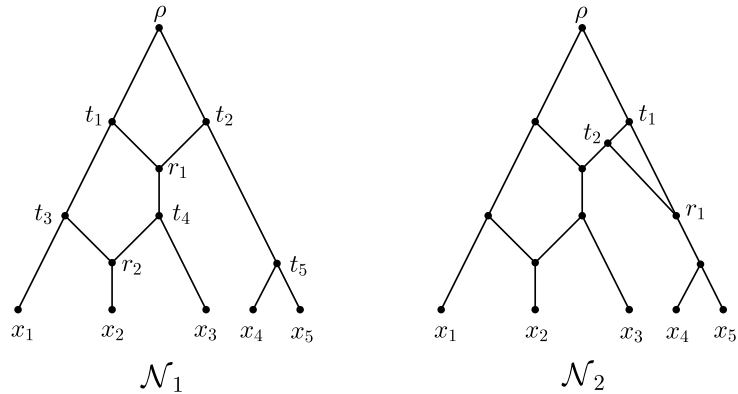}
    \caption{Temporal phylogenetic network $\cN_1$ and non-temporal network $\cN_2$.}
    \label{Fig_Temporal}
\end{figure}

Note that while all rooted phylogenetic trees have a temporal labeling, this is not necessarily the case for general networks. Consider, for example, the network $\cN_2$ depicted in Figure~\ref{Fig_Temporal}. To see why no temporal labeling exists for this network, suppose for the sake of a contradiction that $\cN_2$ has a temporal labeling. Then, condition (T1) implies that $t(r_1) = t(t_1) = t(t_2)$. On the other hand, by condition (T2), we have $t(t_1) < t(t_2)$, a contradiction. Network $\cN_1$ depicted in Figure~\ref{Fig_Temporal}, on the other hand, has a temporal labeling. For instance, it is easily verified that the map $t: V \rightarrow \mathbb{N}_{\geq 0}$ with $t(\rho)=0$, $t(t_1)=t(t_2)=t(r_1)=1$, $t(t_3)=t(t_4)=t(t_5)=t(r_2)=2$, and $t(x_1) = t(x_2) = t(x_3) = t(x_4) = t(x_5) = 3$ satisfies conditions (T1) and (T2).

\begin{figure}[htbp]
    \centering
    \includegraphics[scale=0.15]{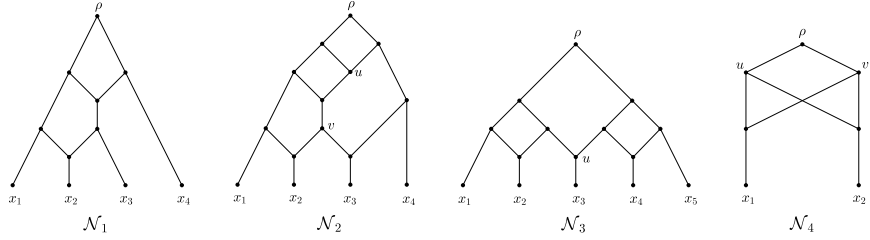}
    \caption{Four phylogenetic networks. Network $\cN_1$ is tree-child and tree-sibling. Network $\cN_2$ is tree-sibling because every reticulation of $\cN_2$ has a sibling that is a tree vertex; however, $\cN_2$ is not a tree-child network because vertices $u$ and $v$ are not parents of tree vertices. Network $\cN_3$ is not tree-sibling (because both siblings of the reticulation vertex $u$ are reticulations) and thus in particular not tree-child, and analogously $\cN_4$ is neither tree-sibling nor tree-child. Moreover, networks $\cN_1, \cN_3$, and $\cN_4$ are stack-free, whereas $\cN_2$ is not (because the reticulation vertex $u$ is the parent of another reticulation vertex). Finally, networks $\cN_1$ and $\cN_3$ are also FU-stable, whereas $\cN_4$ is not (because the two tree vertices $u$ and $v$ have the same set of children).}
    \label{Fig_TreeChild}
\end{figure}

\paragraph{Tree-child networks.}\index{phylogenetic network!tree-child}
A phylogenetic network $\cN$ is a \emph{tree-child} network \citep{Cardona2009} if every non-leaf vertex is the parent of a tree vertex\index{tree vertex} (see Figure~\ref{Fig_TreeChild} for examples of tree-child and non-tree-child phylogenetic networks). Equivalently, for every vertex $v$ of a tree-child network $\cN$, there is a path from $v$ to a leaf that consists only of tree vertices (except the leaf and possibly $v$ itself). This property is also referred to as the \emph{tree-path} property\label{treepathproperty} \citep{Cordue2014}\index{tree-path property}. Note that this also implies that every vertex of a tree-child network is visible\index{visible vertex}. Another equivalent definition of tree-child networks is the following: $\cN$ is a tree-child network if (i) no tree vertex is the parent of two reticulations and (ii) no reticulation is the parent of another reticulation \citep{Semple2015}.

Biologically, tree-child networks represent a scenario where some portion of each non-extant taxon (that may be the ancestor of contemporary taxa) had some descendants that persisted in the environment via mutation instead of hybridization. Such a scenario is more likely in nature than a `non-tree-child' scenario, because although the proportion of extant species with hybrid origin is higher than previously thought, due to pre- and/or post-reproductive barriers, hybridization is still considered relatively uncommon compared to traditional speciation events.

When tree child networks in addition satisfy the time-consistency property, they are referred to as \emph{tree-child time consistent} (TCTC)\index{phylogenetic network!tree-child time-consistent} networks~\citep{Willson2007,Cardona2009a}. Networks in this class are thought to be biologically meaningful~\citep{Willson2007}. In particular, TCTC networks combine the properties of time-consistency  (where (i) tree children temporally exist later than their parents, (ii) hybrid children coexist in time with their parents, and (iii) the parents of a hybrid species also coexist in time) and the tree-child condition, which imposes that every non-extant taxon has some descendants that diverged through mutation alone \citep{cardona2009b}.

\paragraph{Normal and regular networks.} 
A phylogenetic network $\cN=(V,E)$ is a \emph{normal}\index{phylogenetic network!normal} network \citep{Willson2009} if it is tree-child\index{phylogenetic network!tree-child} and has no shortcuts\index{reticulation arc!shortcut}. As an example, both networks depicted in Figure~\ref{Fig_Temporal} are tree-child, but only $\cN_1$ is normal; $\cN_2$ contains a shortcut, arc $(t_1,r_1)$. Note that normal networks `inherit' the biological meaning of tree-child networks. In addition, they have a number of convenient mathematical and computational properties that make them a popular class of networks in mathematical phylogenetics. As an example, normal networks are uniquely characterized by their displayed caterpillar trees on three and four leaves, whereas arbitrary phylogenetic networks cannot be encoded this way (for details and definitions, see Section \ref{subsec_combinatorial_ident}). 

\noindent A phylogenetic network $\cN=(V,E)$ is called a \emph{regular}\index{phylogenetic network!regular} network \citep{Baroni2005} if it satisfies the following three properties (adapted from \cite{Steel2016}):
\begin{enumerate}[(R1)]
    \item If $u, v \in V$ are distinct, then $c_{\cN}(u) \neq c_{\cN}(v)$;
    \item $u \preceq_{\cN} v$ if and only if $c_{\cN}(v) \subseteq c_{\cN}(u)$;
    \item $\cN$ has no redundant arcs/shortcuts.
\end{enumerate}
Notice that by (R1) a regular network cannot contain an edge $(u,v)$ leading from a vertex $u$ of out-degree one to a vertex $v$ of in-degree one. This might seem very restrictive, but stems from the fact that \cite{Baroni2005} introduced this concept for more general networks (so-called `hybrid phylogenies') than the ones considered in this manuscript. It has been argued by \cite{Willson2007} that under an evolutionary model of ``gene aggregation'' (essentially, a perfect phylogeny model assuming binary characters, where each character can mutate only once and is then preserved), only regular networks are meaningful (for further details, see \cite{Willson2007}).

\paragraph{Tree-sibling networks.}\index{phylogenetic network!tree-sibling}
A phylogenetic network $\cN$ is a \emph{tree-sibling} network \citep{Cardona2008} if every reticulation vertex has at least one sibling\index{sibling} that is a tree vertex (cf. Figure~\ref{Fig_TreeChild}). Biologically, for each reticulation event in a tree-sibling network, at least one of the species involved in it also has some descendant through mutation, i.e., through (vertical) descent with modification.
Notice, however, that the class of tree-sibling networks generalizes tree-child networks by allowing non-leaf vertices to be parents solely of reticulation vertices. For instance, the single child of a reticulation vertex can itself be a reticulation vertex as shown in $\cN_2$ in Figure~\ref{Fig_TreeChild}, where the child of the reticulation vertex $u$ is also a reticulation vertex. Biologically, this scenario can be interpreted as the ancestral hybrid taxon $u$ interbreeding with its sibling at the time to produce hybrid taxon $v$ (more precisely, the ancestor of $v$), without persistence of the ancestral taxon $u$ as a distinct lineage.  This might occur, for example, when hybrid taxon $v$ is more fit than $u$, leading $u$ to become extinct.
Finally, note that tree-sibling networks that additionally satisfy the time-consistent property, are referred to as \emph{tree-sibling time consistent}\index{phylogenetic network!tree-sibling time-consistent} (TSTC) networks in the literature \citep{Cardona2008}.

\begin{figure}[htbp]
    \centering
    \includegraphics[scale=0.2]{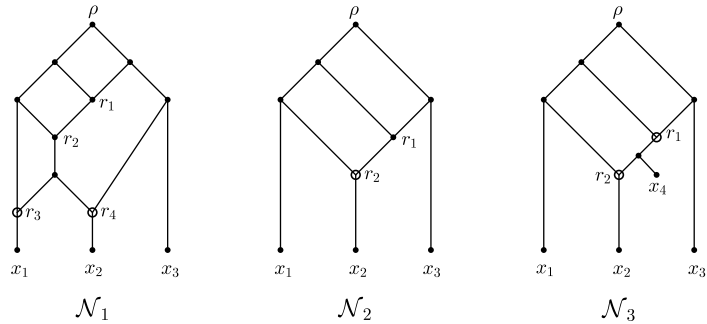}
    \caption{Three phylogenetic networks $\cN_1, \cN_2$, and $\cN_3$ whose visible reticulation vertices are marked as unfilled dots. The network $\cN_1$ is not sink-visible (since $r_2$ is a sink but
    is not visible) and thus in particular not reticulation-visible. The network $\cN_2$ is sink-visible (reticulation $r_2$ is the only sink of $\cN_2$ and is visible) but not reticulation-visible ($r_1$ is not visible). Finally, the network $\cN_3$ is reticulation-visible and thus also sink-visible.}
    \label{Fig_Visibility}
\end{figure}

\paragraph{Reticulation-visible and sink-visible networks.}\index{phylogenetic network!reticulation-visible} \index{phylogenetic network!sink-visible}
A phylogenetic network $\cN$ is called \emph{reticulation-visible} if every reticulation is visible\index{visible vertex} \citep{Gambette2015,vaniersel2010}.
Furthermore, a reticulation vertex is called a \emph{sink}\index{reticulation vertex!sink} if it is the parent of a tree vertex. 
A \emph{sink-visible network} is a phylogenetic network with the property that every sink is visible. Thus, every reticulation-visible network is also sink-visible, whereas the converse is not true. 
For an illustration of these concepts, see Figure~\ref{Fig_Visibility}.

Biologically, the visibility of vertices, in particular reticulation vertices, is relevant for reconstructing phylogenetic networks from genomic data observed at the present. If a vertex is not visible, there might not be strong evolutionary signal for its presence, since evolutionary information passed from the root to the leaves of the network could have simply bypassed this vertex.

\paragraph{Stack-free networks.}\index{phylogenetic network!stack-free}
A phylogenetic network $\cN$ is a \emph{stack-free} network \citep{Semple2018} if $\cN$ has no two reticulations, $u$ and $v$ say, such that $u$ is the parent of $v$; that is, there is no reticulation arc in
which both end-vertices are reticulations. Such a pair of reticulations are called \emph{stack reticulations}\index{reticulation vertex!stack reticulations}. In contrast, two distinct reticulations are called \emph{sibling reticulations}\index{reticulation vertex!sibling reticulations} if they have a common parent. Note that stack-free networks generalize tree-child\index{phylogenetic network!tree-child} networks by allowing sibling reticulations. 
As an example, the networks $\cN_1, \cN_3$ and $\cN_4$ depicted in Figure~\ref{Fig_TreeChild} are stack-free, whereas the network $\cN_2$ in this figure is not since both $u$ and its only child are reticulation vertices.

Biologically, stack-free phylogenetic networks imply that a species resulting from a reticulation event is not directly involved in another reticulation event but rather evolves through vertical descent. Note that stack-free phylogenetic networks are uniquely characterized by the fact that there are two phylogenetic tree embeddings that collectively cover all the arcs of the network (for details see~\cite{Semple2018}). Stack-free phylogenetic networks can thus also be interpreted as being an `amalgamation' of two phylogenetic trees \citep{Semple2018}.

\begin{figure}[htbp]
    \centering
    \includegraphics[scale=0.2]{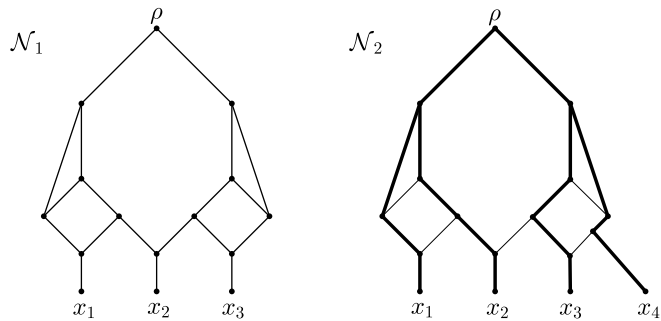}
    \caption{Non-tree-based phylogenetic network $\cN_1$ and tree-based phylogenetic network $\cN_2$. A support tree for $\cN_2$ is highlighted in bold.}
    \label{Fig_Treebased}
\end{figure}

\paragraph{Tree-based networks.}
A phylogenetic network $\cN$ is called \emph{tree-based}\index{phylogenetic network!tree-based} \citep{Francis2015} with \emph{base tree}\index{phylogenetic network!tree-based!base tree} $\cT$ if $\cN$ can be obtained from $\cT$ via the following steps:
\begin{enumerate}[(i)]
    \item Subdivide the arcs of $\cT$, i.e., introduce new vertices of in- and out-degree one, so-called \emph{attachment points}.
    \item Add arcs, so-called \emph{linking arcs}\index{reticulation arc!linking arc}, between pairs of attachment\index{attachment point} points, so that $\cN$ remains binary and acyclic.
    \item Suppress every attachment point that is not incident to a linking arc.
\end{enumerate}
Notice that this procedure allows for parallel edges to be present in a tree-based network (namely, if a linking arc is introduced between two adjacent attachment points). However, these parallel edges can simply be deleted (and the corresponding attachment points suppressed) to be consistent with Definition~\ref{Def_PhylogeneticNetwork}.
Alternatively, a phylogenetic network $\cN$ on $X$ is tree-based if and only if there exists a rooted spanning tree for $\cN$ (that is a rooted tree that contains all vertices and a subset of the arcs of $\cN$) whose leaf set is $X$. Such a spanning tree is also called a \emph{support tree}\index{phylogenetic network!tree-based!support tree} for $\cN$. Examples of a tree-based phylogenetic network and a non-tree-based phylogenetic network are shown in Figure~\ref{Fig_Treebased}.

Moreover, notice that a tree-based network can have different base trees. In fact, there are tree-based networks on $X$, so-called \emph{universal tree-based}\index{phylogenetic network!tree-based!universal} networks \citep{Francis2015, Hayamizu2016, Zhang2016, Bordewich2018}, that have \emph{every} phylogenetic $X$-tree as a base tree. 

Tree-based phylogenetic networks were introduced by \cite{Francis2015} as a way of quantifying the notion of an `underlying tree', i.e., as a way to approach the question of whether a phylogenetic network is merely a phylogenetic tree with some additional horizontal edges, or whether a phylogenetic network has little resemblance to a tree and the concept of an underlying tree should be discarded. Tree-based networks are thus relevant to the continuing debate and discussion in the literature on whether evolution is tree-like with occasional non-tree-like events such as horizontal gene transfer \citep{Daubin2003, Kurland2003}, or whether evolution is inherently network-like and has no tree-like similarities at all \citep{Dagan2006, Doolittle2007, Martin2011, Corel2016}.

\paragraph{LGT networks and species graphs.}
Recall that all arcs directed into a reticulation are referred to as \emph{reticulation arcs}, and are treated symmetrically in the sense that both parents of a reticulation vertex play a symmetrical role. While this is suitable for modelling hybrid speciation\index{hybridization!hybrid speciation} and recombination\index{recombination} events, it is less suitable for modelling LGT\index{lateral gene transfer} (or introgression\index{hybridization!introgression}) events.
In these cases, one of the arcs directed into a reticulation might rather be seen as a `backbone tree arc' instead of a reticulation arc as in Figure~\ref{Fig_Reticulation}(b). This concept is made more precise by the class of \emph{LGT networks} \citep{Cardona2015, Scornavacca2017}. While the concept is related to tree-based networks, the approach is specifically designed to model LGT events and to emphasize the asymmetrical role of the parents of a reticulation.

An \emph{LGT network}\index{phylogenetic network!LGT network}\footnote{We are using the definition from \cite{Scornavacca2017} for binary networks, which is slightly different from the definition originally considered by \cite{Cardona2015} for more general networks.} is a phylogenetic network $\cN=(V,E)$ on $X$ along with a bipartition of $E$ in a set of \emph{principal arcs}\index{tree arc!principal arc} $E_p$ and a set of \emph{secondary arcs}\index{reticulation arc!secondary arc} $E_s$, such that $\cT_0 = (V,E_p)$ is a phylogenetic $X$-tree up to suppression of vertices of in-degree one and out-degree one. In this sense, LGT networks are tree-based with a uniquely distinguished base tree\index{phylogenetic network!tree-based!base tree}. Principal arcs explicitly model the primary (tree-like) line of evolution, and secondary arcs model the LGT events. As an example, for the tree-based network $\cN_2$ depicted in Figure~\ref{Fig_Treebased}, the bold arcs can be seen as corresponding to the primary tree-like evolution of taxa $x_1, \ldots, x_4$, whereas the thin arcs can be interpreted as events of LGT.

Note that the distinction of `principal' and `secondary' arcs also arises in the context of network inference methods assuming a coalescent process when not only the network topology but also numerical parameters for the inheritance probabilities\index{inheritance probability} (p.~\pageref{inheritanceprob}) associated with reticulation arcs are estimated. For example, in PhyloNetworks \citep{solis-lemus2016,solis-lemus2017}, the reticulation arc that is assigned the lower probability is referred to as the `minor hybrid edge', and the other  reticulation arc is called the `major hybrid edge'. It is then also possible to extract the `major tree', i.e., the tree obtained from deleting the minor hybrid edge at each reticulation \citep{solis-lemus2016} (see also \cite{Jiao2021}).
 
Another class of networks that explicitly distinguishes between the primary and secondary line of evolution are \emph{species graphs}\index{phylogenetic network!species graphs} as defined by \cite{Gorecki2004}. More restrictive than LGT networks, these are composed of a principal tree and a set of secondary arcs, where the secondary arcs must satisfy a set of more restrictive conditions than in LGT networks. For instance, the resulting network must be time consistent, and the head and source of a secondary arc cannot be connected by a path in the principal tree. We refer the reader to \cite{Gorecki2004} for further details.

\begin{figure}[htbp]
    \centering
    \includegraphics[scale=0.175]{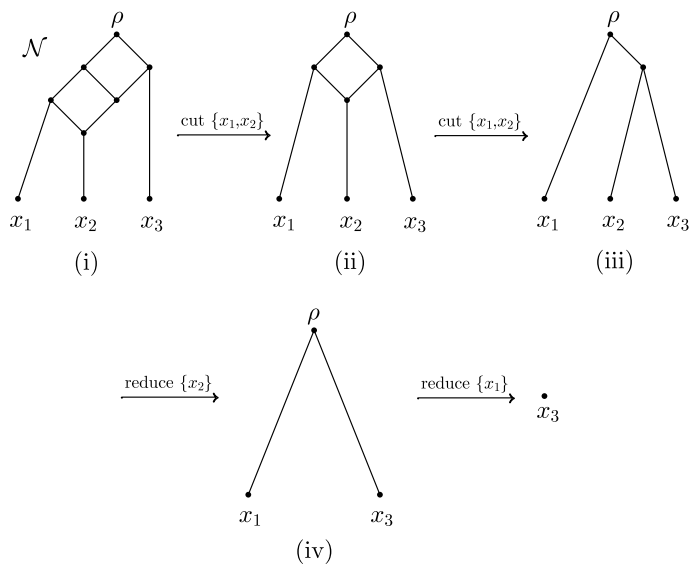}
    \caption{Orchard network $\cN$ on $X=\{x_1,x_2,x_3\}$ and a complete sequence of cherry reductions that reduces it to a single vertex.}
    \label{Fig_Orchard}
\end{figure}

\paragraph{Orchard or cherry-picking networks.}
Let $\cN$ be a phylogenetic network on $X$, and let $x,y \in X$ be two distinct leaves of $\cN$. Let $p_x$ and $p_y$ denote the parents of $x$ and $y$, respectively. If $p_x=p_y$, the pair $\{x,y\}$ is called a \emph{cherry}\index{cherry} of $\cN$. Furthermore, if one of the parents, say $p_y$, is a reticulation and there is an edge $(p_x,p_y)$ in $\cN$, then $\{x,y\}$ is called a \emph{reticulated cherry}\index{cherry!reticulated cherry} of $\cN$ with \emph{reticulation leaf} $y$. As an example, consider Figure~\ref{Fig_Orchard}. Here, the phylogenetic network $\cN$ in panel (i) contains a reticulated cherry, namely $\{x_1,x_2\}$, and $x_2$ is the reticulation leaf. Moreover, the phylogenetic network in panel (iii) contains a cherry, namely the pair $\{x_2,x_3\}$.

Now, there are two \emph{cherry reductions}\index{cherry!cherry reduction} \citep{Erdos2019} (see also \cite{Janssen2021}) associated with cherries and reticulated cherries:
\begin{itemize}
    \item If $\{x,y\}$ is a cherry\index{cherry} of $\cN$, \emph{reducing $y$} is the operation of deleting $y$ and suppressing the resulting vertex of in-degree one and out-degree one. If the parent of $x$ and $y$ is the root of $\cN$, then reducing $y$ consists of deleting $y$ as well as the root of $\cN$, resulting in an isolated vertex $x$. 
    \item If $\{x,y\}$ is a reticulated cherry\index{cherry!reticulated cherry} of $\cN$ in which $y$ is the reticulation leaf, \emph{cutting} $\{x,y\}$ is the operation of deleting the reticulation edge $(p_x,p_y)$, and suppressing the resulting two vertices of in-degree one and out-degree one. 
\end{itemize}
Now, if a phylogenetic network $\cN$ can be reduced to a single vertex by a sequence of cherry reductions, it is called an \emph{orchard} network\index{phylogenetic network!orchard} \citep{Erdos2019} or a \emph{cherry-picking} network\index{phylogenetic network!cherry-picking} \citep{Janssen2021} (note that \cite{Erdos2019} and \cite{Janssen2021} independently introduced this class of networks). An example of an orchard network and a possible complete sequence of cherry reductions\footnote{Note that there might be several complete sequences of cherry reductions that reduce an orchard network to a single leaf and the order in which cherry reductions are performed does not matter \citep[Proposition~4.1]{Erdos2019}.} that reduces it are depicted in Figure~\ref{Fig_Orchard}.  

While orchard networks were originally introduced without any biological justification, \citet{vanIersel2021} recently showed that they are characterized by admitting a \emph{HGT-consistent labelling}. Intuitively, this means that orchard networks \enquote{are consistent with an evolutionary history in time in which reticulate events represent instantaneous (horizontal) transfers such as LGT events} \citep{vanIersel2021}. In this sense, orchard networks can be seen as trees with additional \emph{horizontal} arcs, making them biologically highly relevant. Note that orchard networks are closely related to tree-based networks\index{phylogenetic network!tree-based}, except that the additional arcs in tree-based networks do not need to be horizontal. In particular, every orchard network is tree-based (e.g., \cite{Huber2019}). They are also closely related to LGT networks\index{phylogenetic network!LGT network}, with the difference that orchard networks do not (necessarily) specify which arcs are secondary arcs, i.e., LGT arcs \citep{vanIersel2021}.

Orchard networks also exhibit certain desirable mathematical properties. As an example, they are uniquely characterized by their induced subnetworks on three leaves and can also be encoded by certain sets of paths in the network (for details and formal definitions, see Section~\ref{subsec_combinatorial_ident}).

\paragraph{Galled trees, galled networks, and level-\texorpdfstring{$k$}{k} networks.}
Let $\cN$ be a phylogenetic network. A \emph{reticulation cycle}\index{reticulation cycle} of $\cN$ is a pair of directed paths with a common start vertex and common end vertex that are vertex-disjoint otherwise (note that the common start vertex is thus necessarily a tree vertex and the common end vertex is necessarily a reticulation). Note that every reticulation of $\cN$ lies on at least one reticulation cycle, and this vertex could either be the end vertex of the two paths defining that cycle or an intermediate vertex on one of them. In fact, if a network contains several reticulation cycles, they might be dispersed or they might be more or less interwoven. 

\begin{figure}[htbp]
    \centering
    \includegraphics[scale=0.15]{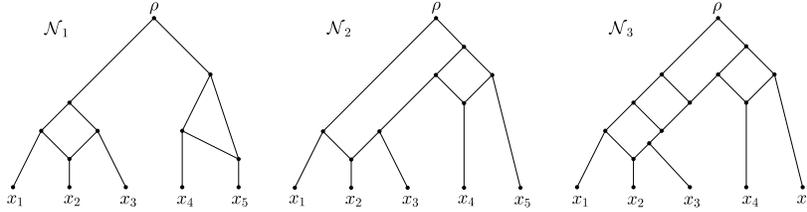}
    \caption{Galled tree $\cN_1$, galled network $\cN_2$ (note that the reticulation cycles of $\cN_2$ share tree vertices but no reticulation vertices), and a phylogenetic network $\cN_3$ that is neither a galled tree nor a galled network. Note that $\cN_1$ is a level-$1$ network, $\cN_2$ is a level-$2$ network, and $\cN_3$ is a level-$4$ network.}
    \label{Fig_Level}
\end{figure}

Now, if the reticulations cycles of a network $\cN$ are all vertex-disjoint (i.e., the reticulation cycles are isolated from each other), $\cN$ is said to be a \emph{galled tree}\index{phylogenetic network!galled tree} \citep{Gusfield2003}\footnote{Note that the property of vertex-disjoint reticulation cycles was first discussed by \cite{Wang2001}, and later called the ``gall property'' by \cite{Gusfield2003}.} In this case, no two reticulation vertices can be contained in a common reticulation cycle, and reticulation events can be seen as independent from each other. An example of a galled tree is the phylogenetic network $\cN_1$ depicted in Figure~\ref{Fig_Level}. Note that if a galled tree, or more generally a phylogenetic network, only contains one reticulation cycle\index{reticulation cycle}, it is also called a \emph{unicyclic} network.

A generalization of galled trees are \emph{galled networks}\index{phylogenetic network!galled network} \citep{Huson2007}, where reticulation cycles may share tree vertices (and thus they may overlap on tree arcs). The previous assumption for galled trees is thus relaxed in the sense that the only constraint is that every reticulation vertex is contained in a reticulation cycle formed by paths composed exclusively by tree vertices. As an example, the phylogenetic network $\cN_2$ depicted in Figure~\ref{Fig_Level} is a galled network but not a galled tree. 

A slightly more general notion quantifying whether reticulations are widely separated or highly interwoven is the `level' of a network \citep{Choy2005}. To define this, let a \emph{biconnected component} or \emph{block} of a phylogenetic network $\cN$ be a subnetwork $\cN'$ of $\cN$ such that (a) $\cN'$ remains connected if any one of its vertices (together with its incident arcs) is deleted; and (b) $\cN'$ is maximal with respect to property (a). A phylogenetic network $\cN$ is said to be a \emph{level-$k$ network}\index{phylogenetic network!level-$k$} \citep{Choy2005} if every biconnected component of $\cN$ has at most $k$ reticulation vertices. Note that a level-$0$ network is a phylogenetic tree, and a level-$1$ network\index{phylogenetic network!level-$k$!level-$1$} is a galled tree as defined above. An illustration of this concept is given in Figure~\ref{Fig_Level}, where $\cN_1$ is a level-$1$ network, $\cN_2$ is a level-$2$ network, and $\cN_3$ is a level-$4$ network.

Biologically, the level of a network and related concepts provide another way (next to tree-basedness) to assess whether a network can be viewed as mainly tree-like with some local and dispersed reticulations, or whether it is highly tangled and interwoven with little resemblance to a tree-like structure.  

Mathematically, restricting the level of a network often leads to more tractable problems. For example, most studies on the statistical identifiability of phylogenetic networks to date have focused on level-$1$ networks, and similarly many positive results concerning the combinatorial identifiability of phylogenetic networks are restricted to networks of a small level (see Section \ref{Sec_Identifiability}, where we discuss identifiability questions in more detail). In addition, restricting the level of a network is currently a common way to obtain scalable network inference methods. As an example, both PhyloNetworks \citep{solis-lemus2016,solis-lemus2017} and NANUQ \citep{Allman2019} consider only level-$1$ networks (note that we discuss  this in more detail in Section~\ref{Sec_Scalability}).

\subsection{Classes of rooted binary phylogenetic networks not yet linked 
to biology} \label{Subsec_notlinked}
In the following, we present additional classes of rooted binary phylogenetic networks that have not been linked to a particular biological process yet and are mostly of mathematical or algorithmic relevance to date.

\paragraph{FU-stable networks.} \index{phylogenetic network!FU-stable}
A phylogenetic network $\cN$ is called an \emph{FU-stable network}\footnote{Note that \cite{Huber2016a} originally simply called such networks \emph{stable}. However, as the word ``stable'' is used in various contexts in the phylogenetic networks literature, we refer to FU-stable networks to avoid any ambiguity.} \citep{Huber2016a} if it is (i) stack-free and (ii) has the property that any two distinct tree vertices of $\cN$ have distinct sets of children \citep[Theorem 1]{Huber2016a}. As an example, while the networks $\cN_1, \cN_3$, and $\cN_4$ depicted in Figure~\ref{Fig_TreeChild} are all stack-free, only $\cN_1$ and $\cN_3$ are FU-stable. Network $\cN_4$ is not FU-stable because the two distinct tree vertices $u$ and $v$ have identical sets of children.

Note that the class of FU-stable networks arises in the context of `folding and unfolding' phylogenetic networks that we briefly mentioned in Section~\ref{Sec_NetworkDef}. More precisely, \cite{Huber2016a} call a phylogenetic network $\cN$ FU-stable if $F(U(\cN))$ is isomorphic to $\cN$, where -- informally speaking -- $U(\cN)$ is the MUL-tree\index{MUL-tree} obtained from `unfolding' $\cN$ and $F(U(\cN))$ is the phylogenetic network obtained from `folding up' $U(\cN)$ (for a more rigorous definition and additional details, see \cite{Huber2016a}). In particular, the definition given here is rather a characterization of the class of FU-stable networks, whereas the original definition of \cite{Huber2016a} is motivated by the folding--unfolding process applied to a phylogenetic network.

In addition, note that the process of `unfolding' a phylogenetic network $\cN$ into a MUL-tree is used in PhyloNet \citep{than2008,wen2018a}, a popular network inference tool, as one way of calculating the probability of a gene tree given a species network (for details, see \cite{Yu2012,yuetal2014}).

\paragraph{\texorpdfstring{$k$}{k}-nested networks.}
In order to define the class of $k$-nested networks introduced by \cite{Jansson2004}, we need some additional terminology (adapted from \cite{Jansson2004}). First, let $\cN=(V,E)$ be a phylogenetic network and let $r$ be a reticulation vertex. Then, every tree vertex $t$ that is an ancestor of $r$ such that $r$ can be reached using two disjoint directed paths starting at the children of $t$ is called a \emph{split node of $r$}\index{tree vertex!split node}. If $t$ is a split node for $r$, any directed path from $t$ to $r$ is a \emph{merge path of $r$}, and any path from a child of $t$ to a parent of $r$ is a \emph{clipped merge path of $r$}. \cite{Jansson2004} call a phylogenetic network \emph{nested}\index{phylogenetic network!nested} if for every two merge paths $P_1,P_2$ of two distinct reticulation vertices $r_1$ and $r_2$, $P_1$ and $P_2$ share a common arc if and only if one of these paths is a subpath of the other. Moreover, for a nested phylogenetic network $\cN=(V,E)$ and a vertex $v \in V$, the \emph{nesting depth of $v$}, denoted by $d(v)$, is the number of reticulation vertices in $\cN$ that have a clipped merge path traversing $v$. Finally, the \emph{nesting depth of $\cN$}, denoted by $d(\cN)$, is the maximum value of $d(v)$ over all $v \in V$, and $\cN$ is called \emph{$k$-nested} if it is nested with nesting depth at most $k$.\index{phylogenetic network!nested!$k$-nested}
Note that $d(\cN) \leq 1$ if and only if $\cN$ is a galled tree\index{phylogenetic network!galled tree}. Moreover, note that if $\cN$ is a nested phylogenetic network with nesting depth $d$, then we have for the level of $\cN$, say $k$, that $k \geq d$, i.e., the nesting depth is a lower bound for the level of $\cN$ \citep{Jansson2004}. 

\begin{figure}[htbp]
    \centering
    \includegraphics[scale=0.15]{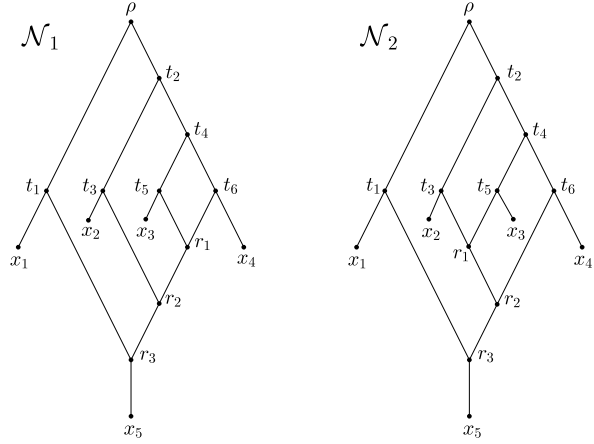}
    \caption{Nested phylogenetic network $\cN_1$ with nesting depth $3$, and phylogenetic network $\cN_2$ that is not nested. To see that $\cN_2$ is not nested, notice that $t_2$ is a split node for $r_2$ and $\rho$ is a split node for $r_3$. However, the two merge paths $P=(t_2, t_4, t_5, r_1, r_2)$ and $P'=(\rho, t_2, t_4, t_6,r_2,r_3)$ share an arc, namely the arc $(t_2,t_4)$, but $P$ is not a subpath of $P'$ and vice versa.}
    \label{Fig_Nested}
\end{figure}

As an example, the phylogenetic network $\cN_1$ depicted in Figure~\ref{Fig_Nested} is nested with nesting depth $3$, whereas the phylogenetic network $\cN_2$ depicted in the same figure is not nested. In case of $\cN_1$, $\rho$ is the unique split node\index{tree vertex!split node} for $r_3$, $t_2$ is the unique split node for $r_2$, and $t_4$ is the unique split node for $r_1$. 
The collection of merge paths for $\cN_1$ is given by $\{\tuple{\rho, t_1, r_3}, \tuple{\rho, t_2, t_3, r_2, r_3}, \tuple{\rho,t_2,t_4,t_5,r_1,r_2,r_3}, \tuple{\rho,t_2,t_4,t_6,r_1,r_2,r_3}, \tuple{t_2,t_3,r_2}, \tuple{t_2,t_4,t_5,r_1,r_2}, \tuple{t_2,t_4,t_6,r_1,r_2}, \tuple{t_4,t_5,r_1}, \tuple{t_4,t_6,r_1}\}$ and it can be easily checked that any two merge paths $P_1,P_2$ for two distinct reticulation vertices share a common arc if and only if one of the paths is a subpath of the other. 
Moreover, to see that the nesting depth of $\cN_1$ is equal to 3, notice that vertices $t_5$ and $t_6$ have a nesting depth of $d(t_5)=d(t_6)=3$ (and as $\cN_1$ contains precisely three reticulations this is maximal).

\paragraph{\texorpdfstring{$k$}{k}-reticulated networks.} A phylogenetic network $\cN$ is called \emph{$k$-reticulated}\index{phylogenetic network!$k$-reticulated} \citep{Vu2013} if for any vertex $v$ of in-degree at most one (i.e., for any tree vertex or the root of $\cN$), there are at most $k$ reticulation vertices that can be reached from $v$ by at least two directed vertex-disjoint paths from $v$ (i.e., paths whose only shared vertices are the start vertex $v$ and the end vertex). Note that every level-$k$ network is also a $k$-reticulated network, but some level-$k'$ networks are in fact $k$-reticulated networks with $k < k'$ \citep{Vu2013}. As an example, the network $\cN_1$ depicted in Figure~\ref{Fig_Level} is a level-$1$ network and is $1$-reticulated. However, the network $\cN_2$ depicted in this figure is a level-$2$ network that is $1$-reticulated.

\paragraph{Spread-\texorpdfstring{$k$}{k} networks.}
The class of \emph{spread-$k$} networks was introduced by \cite{Asano2010} and was motivated by the aim of developing an efficient representation of the clusters of a phylogenetic networks to make the computation of the Robinson-Foulds distance (\cite{Robinson1981}; see also \cite{Cardona2009}) between networks more efficient. Recall that given a phylogenetic network $\cN=(V,E)$ on $X$ and a vertex $v\in V$, the cluster $c_{\cN}(v)$ associated with $v$ is the subset of the leaf set $X$ that can be reached from $v$ via a directed path, i.e., $c_{\cN}(v) = \{x \in X: v \preceq_{\cN} x\}$. Now, following \cite{Asano2010}, a \emph{leaf numbering function} is a bijection from the leaf set $X$ to the set $\{1,\ldots,n\}$. Moreover, for any leaf numbering function $f$ and a vertex $v \in V$, the \emph{characteristic vector for $v$ under $f$}, denoted $C_f[v]$, is a bit vector (i.e., a vector containing zeros and ones) of length $n$ such that for any $i \in \{1,\ldots,n\}$, the $i^{th}$ bit equals 1 if and only if $f^{-1}(i)$ is contained in the cluster $c_{\cN}(v)$ associated with $v$. Note that $C_f[\rho]=11 \dots 11$ for the root $\rho$ of $\cN$ and that $C_f[x]$ contains precisely one 1 for each leaf $x \in X$. Furthermore, a maximal consecutive sequence of ones in a bit vector is called an \emph{interval}. Now, given a leaf numbering function $f$ and a vertex $v \in V$, let $I_f(v)$ denote the number of intervals in $C_f[v]$. Then, the \emph{spread of $f$} is defined as  $I_f 
= \max\limits_{v \in V} I_f(v)$ and the \emph{minimum spread of $\cN$} is the minimum value of $I_f$ taken over all possible leaf numbering functions $f$. Finally, a phylogenetic network $\cN$ is a spread-$k$ network if its minimum spread is at most $k$.\index{phylogenetic network!spread-$k$} 
Note that a level-$k$ network has minimum spread at most $k+1$ \citep{Asano2010}. 

As an example, consider the phylogenetic network $\cN_1$ depicted in Figure~\ref{Fig_Nested}. The minimum spread of $\cN_1$ is 2, and is, for example, achieved by the leaf numbering function $f:\{x_1, \ldots, x_5\} \rightarrow \{1,\ldots,5\}$ with $f(x_i)=i$ for $i=1, \ldots,5$. Here, $C_f[t_1]=(1,0,0,0,1)$, $C_f[t_2]=(0,1,1,1,1)$, $C_f[t_3]=(0,1,0,0,1)$, $C_f[t_4]=(0,0,1,1,1)$, $C_f[t_5]=(0,0,1,0,1)$, $C_f[t_6]=(0,0,0,1,1)$, and $C_f[r_1]=C_f[r_2]=C_f[r_3]=(0,0,0,0,1)$ (as well as $C_f[\rho]=(1,1,1,1,1)$, $C_f[x_1]=(1,0,0,0,0)$, $C_f[x_2]=(0,1,0,0,0)$, $C_f[x_3]=(0,0,1,0,0)$, $C_f[x_4]=(0,0,0,1,0)$, and $C_f[x_5]=(0,0,0,0,1)$). In particular, $I_f[t_1]=I_f[t_3]=I_f[t_5]=2$ (and $I_f[v]=1$ for all other vertices $v$), which leads to $I_f=2$ (as it can easily be checked that there exists no leaf numbering, say $f'$, with $I_{f'} < 2$).

\paragraph{Nearly stable, genetically stable, nearly tree-child, and stable-child networks.}
A phylogenetic network $\cN$ is \emph{nearly stable}\index{phylogenetic network!nearly stable} \citep{Gambette2015} if for every vertex, either the vertex or its parents are visible\index{visible vertex}. In other words, for each arc $(u,v)$ of $\cN$, either $u$ or $v$ (or both) are visible.

$\cN$ is called \emph{genetically stable}\index{phylogenetic network!genetically stable}  \citep{Gambette2016} if every reticulation vertex is visible\index{visible vertex} and has at least one visible parent. This means that a reticulation vertex `inherits' the visibility property from one of its parents. 

Moreover, $\cN$ is called \emph{nearly tree-child}\index{phylogenetic network!nearly tree-child} \citep{Gambette2016} if the following two conditions hold: (i) $\cN$ is reticulation-visible\index{phylogenetic network!reticulation-visible} and (ii) every reticulation of $\cN$ has the tree-path property\index{tree-path property} (as introduced on p.~\pageref{treepathproperty}).

Finally, \emph{stable-child} \index{phylogenetic network!stable-child} networks \citep{Gunawan2015} are phylogenetic networks in which every vertex has at least one visible\index{visible vertex} child. 

\begin{figure}[htbp]
    \centering
    \includegraphics[scale=0.2]{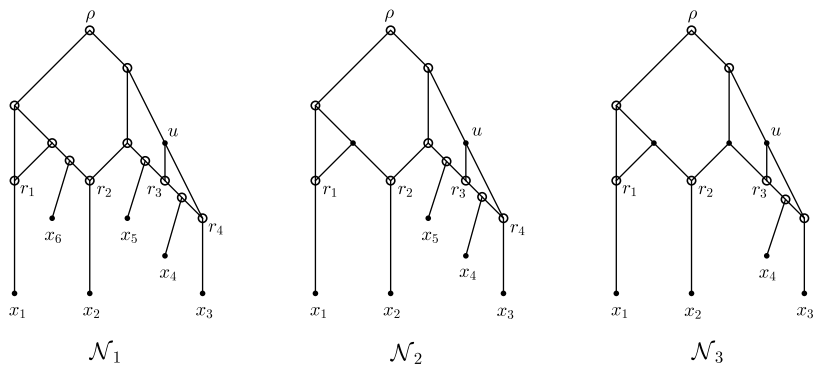}
    \caption{Three reticulation-visible but non-tree-child phylogenetic networks $\cN_1$, $\cN_2$ and $\cN_3$ for which the visible vertices (except for leaves) are depicted as unfilled dots. The network $\cN_1$ is nearly stable, genetically stable, nearly tree-child, and stable-child. The network $\cN_2$ is also nearly stable, genetically stable, and stable-child, but not nearly tree-child (because reticulation $r_1$ does not have a parent that is connected to a leaf of $\cN_2$ by a tree-path). Finally, the network $\cN_3$ is nearly stable, but neither genetically stable (as the reticulation $r_2$ does not have a visible parent), nor nearly tree-child (as reticulations $r_1$ and $r_2$ do not have a parent that is connected to a leaf of $\cN_3$ by a tree-path), nor stable-child (since the parent of $u$ does not have a visible child).}
    \label{Fig_NearlyConcepts}
\end{figure}

Examples for these types of networks are depicted in Figure~\ref{Fig_NearlyConcepts}.

\paragraph{Valid networks.}
The class of \emph{valid} networks was very recently introduced by \cite{Murakami2019} and requires some additional terminology (all subsequent definitions are adapted from \cite{Murakami2019}). 

First, given a directed acyclic graph $G=(V,E)$ (possibly containing some labeled vertices) \emph{cleaning up} $G$ is the act of applying the following operations until none is applicable:
\begin{enumerate}
    \item delete an unlabeled vertex of out-degree zero;
    \item suppress a vertex of in-degree one and out-degree one;
    \item replace a pair of parallel arcs by a single arc.
\end{enumerate}
Second, given a phylogenetic network $\cN$, the deletion of a reticulation arc is \emph{valid} if the resulting subnetwork, after cleaning up, contains exactly two vertices and three arcs fewer than the original network, that is, only the reticulation arc is deleted and its endpoints are suppressed. A reticulation arc is called \emph{valid}\index{reticulation arc!valid} if its deletion is valid in this sense; otherwise it is called \emph{invalid}.

\begin{figure}[htbp]
    \centering
    \includegraphics[scale=0.2]{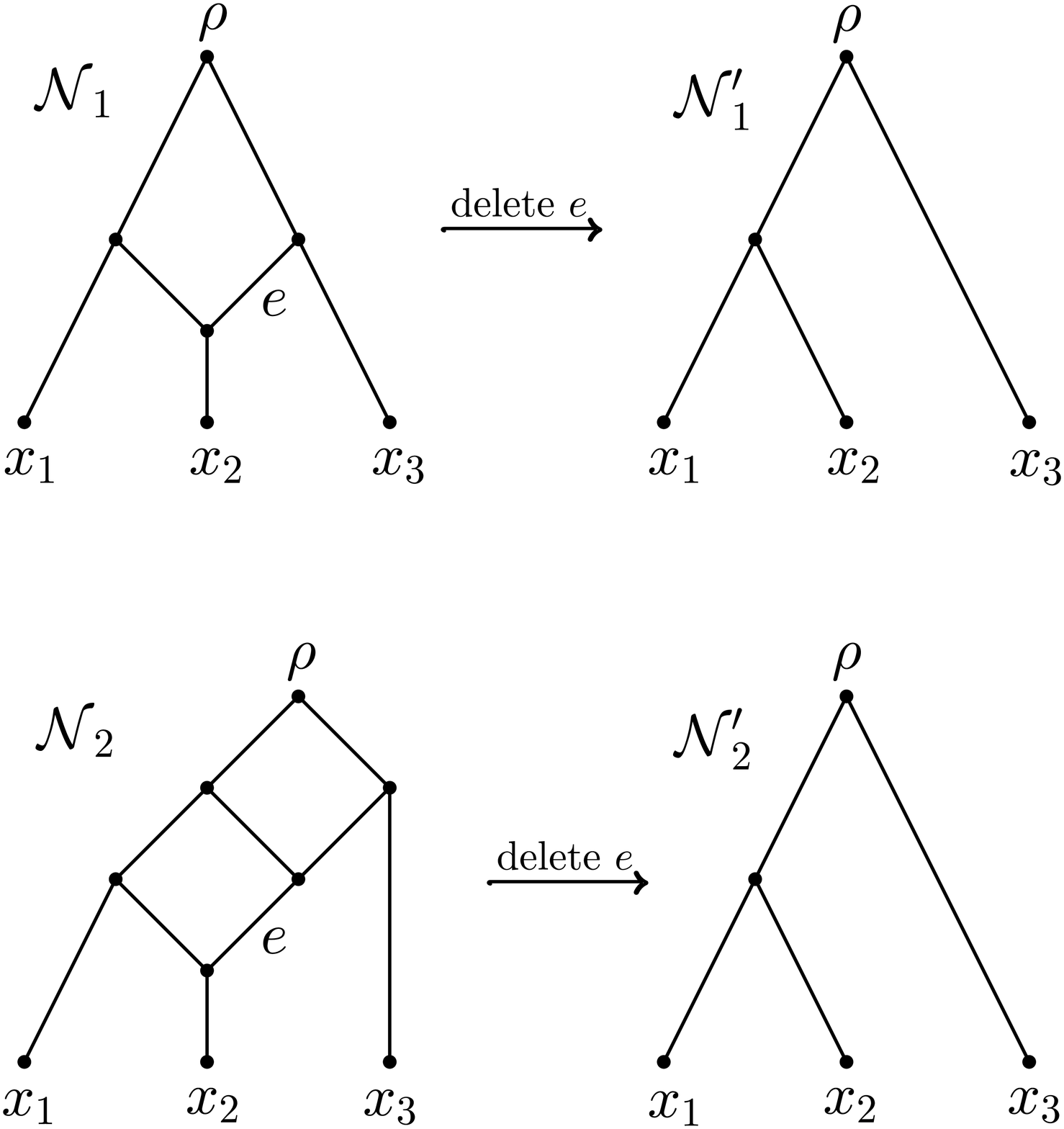}
    \caption{Valid phylogenetic network $\cN_1$ and non-valid phylogenetic network $\cN_2$. To see that $\cN_1$ is valid note that the deletion of the reticulation arc $e$ is valid since the resulting network $\cN_1$ contains precisely two vertices and three fewer arcs than $\cN_1$ after cleaning up (by symmetry the second reticulation arc of $\cN_1$ is also valid). The network $\cN_2$ on the other hand is not valid since the deletion of the reticulation arc $e$ is not valid. More precisely, the deletion of $e$ results in a network $\cN_2'$ with four fewer vertices and six fewer arcs than the original network (Figure adapted from \cite[Fig.~3]{Murakami2019}).}
    \label{Fig_Valid}
\end{figure}

Based on this, a phylogenetic network $\cN$ is called a \emph{valid} network\index{phylogenetic network!valid} \citep{Murakami2019} if all its reticulation arcs are valid. An illustration of this concept can be found in Figure~\ref{Fig_Valid}.

\subsection{Relationship among network classes within rooted binary phylogenetic networks.} 
\begin{figure}[htbp]
    \centering
    \includegraphics[scale=0.55]{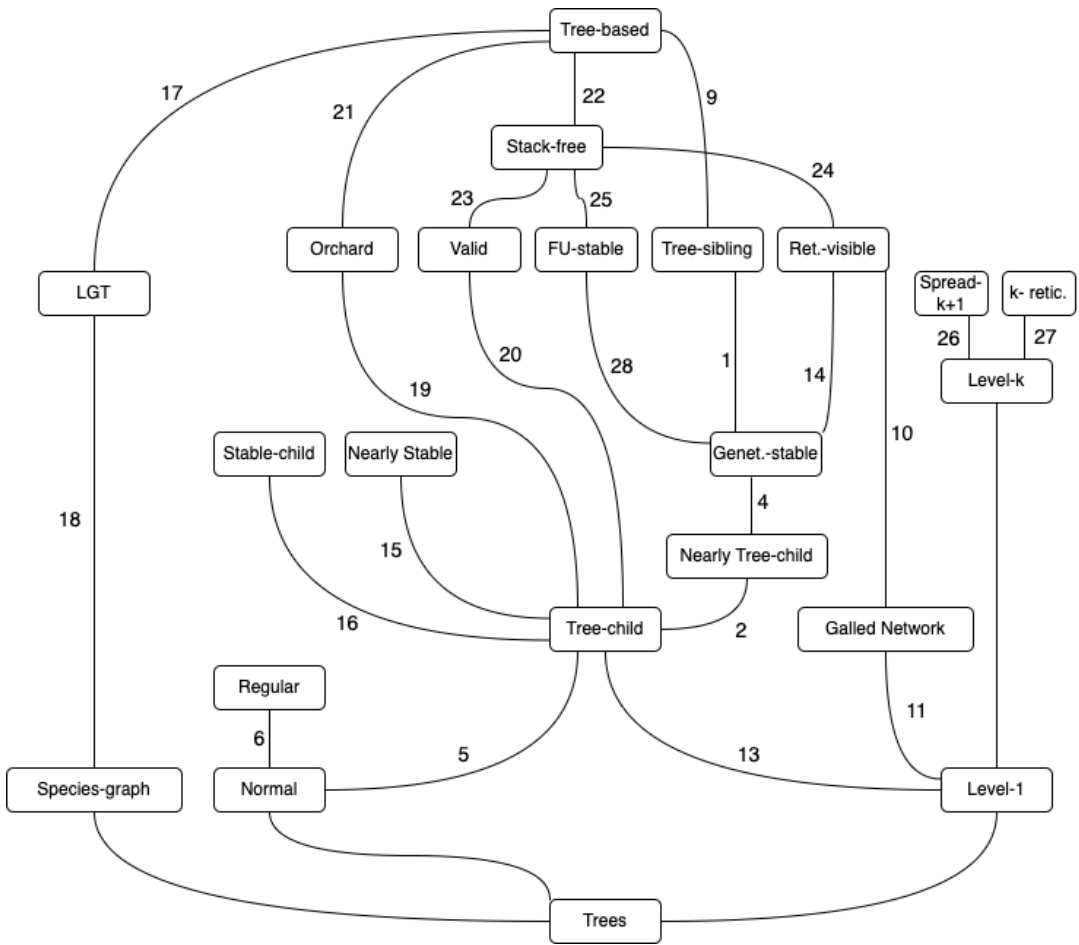}
     \caption{Diagram illustrating the inclusion relationships among network classes within rooted binary networks as given in Table~\ref{tab:Inclusion}.}
    \label{fig:inclusions}
\end{figure}

While we have introduced more than 20 different classes of rooted binary phylogenetic networks in the two preceding sections, some of them are closely related in the sense that one class is a proper subset of another class. We summarize these inclusion relationships among network classes within rooted binary phylogenetic networks in Table~\ref{tab:Inclusion} and additionally visualize them in Figure~\ref{fig:inclusions}. An excellent overview of some of these inclusion-relationships among phylogenetic network classes can also be found on the website \enquote{ISIPhyNC} (\url{https://phylnet.univ-mlv.fr/isiphync/index.php}) \citep{Gambette2018a} mentioned in the introduction.

\begin{longtable}{lll}
\caption{Inclusion relationships among network classes within rooted binary phylogenetic networks. Note that we omit trivial inclusions such as ``level-$k$ is level-$(k+1)$'' for the classes of level-$k$, spread-$k$, $k$-reticulated, and $k$-nested networks.} \\
\toprule
\label{tab:Inclusion}
\centering
\setlength{\tabcolsep}{5mm}
\#   & Inclusion & Reference \\
\midrule
\endfirsthead
\multicolumn{3}{c}%
{\textbf{Table \ref{tab:Inclusion}}: \textit{Continued from previous page}} \\
\toprule
\#   & Inclusion & Reference \\
\midrule
\endhead
1 & \bigcell{l}{genetically stable \\ is tree-sibling} & \citet[Prop.~3.2(1)]{Gambette2018}  \\ \addlinespace
2 & \bigcell{l}{tree-child \\ is nearly tree-child} & definition \\ \addlinespace
3 & \bigcell{l}{reticulation-visible \\ is tree-based} & \citet[Lemma~1]{Gambette2015}  \\ \addlinespace
4 & \bigcell{l}{nearly tree-child \\ is genetically stable} & definition  \\\addlinespace
5 & normal is tree-child & definition \\\addlinespace
6 & \bigcell{l}{compressed\footnotemark normal \\ is regular} & \citet[Theorem 3~.4]{Willson2009}  \\\addlinespace
7 & \bigcell{l}{tree-child \\ is tree-sibing} & \bigcell{l}{\citet[p.~2]{Cardona2008} \\ (see also \citet[Lemma~10.8]{Steel2016})} \\  \addlinespace
8 & \bigcell{l}{tree-child is \\ reticulation-visible} & definition  \\ \addlinespace
9 & \bigcell{l}{tree-sibling \\ is tree-based} & \citet[Corollary~2]{Francis2015}\\ \addlinespace
10 & \bigcell{l}{a galled network is \\ reticulation-visible} & \citet[Lemma~6.11.14]{Huson2011} \\ \addlinespace
11 & \bigcell{l}{a galled tree is \\  a galled network} & definition  \\ \addlinespace
12 & \bigcell{l}{galled trees are \\ tree-sibling} & \citet[Lemma~6.11.16]{Huson2011}\\ \addlinespace
13 & \bigcell{l}{galled trees are \\ tree-child} & \citet[Lemma~6.11.11]{Huson2011} \\ \addlinespace
14 & \bigcell{l}{genetically stable is \\ reticulation-visible} & definition \\ \addlinespace
15 & \bigcell{l}{tree-child is \\ nearly stable} & definition \\ \addlinespace
16 & \bigcell{l}{tree child is \\ stable-child} & definition  \\ \addlinespace
17 & LGT is tree-based & definition \\ \addlinespace
18 & species graph is LGT & \citet[Section~2.3]{Pons2016} \\ \addlinespace
19 & \bigcell{l}{tree-child \\ is orchard} & \citet[p.~138]{Janssen2021} \\ \addlinespace
20 & \bigcell{l}{tree-child \\ is valid} & \citet[Lemma~2]{Murakami2019} \\ \addlinespace
21 & \bigcell{l}{orchard is \\ tree-based} & \citet[Prop.~2]{Huber2019}\\ \addlinespace
22 & \bigcell{l}{stack-free \\ is tree-based} & \citet{Semple2018,Zhang2016} \\ \addlinespace
23 & \bigcell{l}{valid is \\ stack-free} & definition \\ \addlinespace
24 & \bigcell{l}{reticulation-visible \\ is stack-free} & \citet[Theorem~2.1]{Semple2018} \\ \addlinespace
25 & \bigcell{l}{FU-stable is \\ stack-free} & definition \\ \addlinespace
26 & \bigcell{l}{level-$k$ is \\ spread-$(k+1)$} & \citet[Lemma 5]{Asano2010} \\ \addlinespace
27 & \bigcell{l}{level-$k$ is \\ $k$-reticulated} & \citet{Vu2013} \\ \addlinespace
28 & \bigcell{l}{genetically stable is \\ FU-stable} & \bigcell{l}{\citet[Corollary~1]{Huber2016a} \\ (using inclusions 1, 14, 24)}  \\ 
\bottomrule
\footnotetext[7]{Compressed means that every arc $(u,v)$ leading from a vertex $u$ of out-degree one to a vertex $v$ of in-degree one is contracted.}
\end{longtable}

\section{Mathematical and computational challenges in estimating phylogenetic networks}\label{Sec_ScalabilityIdentifiability}

In the previous section, we reviewed several classes of rooted binary phylogenetic networks used in the literature. We saw that some of them have direct biological relevance, whereas others lack an immediate biological interpretation. In the following, we discuss two further aspects that need to be taken into account when estimating phylogenetic networks in an empirical setting, namely scalability and identifiability. In particular, we describe how imposing structural constraints on the networks under consideration, i.e., considering specific subclasses instead of arbitrary phylogenetic networks, can help to address challenges related to scalability and identifiability in network inference and estimation. We remark that both the development of scalable methods in network inference and the analysis of network identifiability are currently very active areas of research in phylogenetics. Our aim is thus to convey and summarize the central ideas and results rather than to recapitulate the technical details.

\subsection{Scalability} \label{Sec_Scalability}
In this section, we consider the task of estimating a phylogenetic network given data for a collection of taxa. This leads to two distinct challenges related to scalability. First, we must evaluate the fit of a specified network to a given data set under a chosen model or optimality criterion. Second, we must search the space of possible networks for those that are optimal under the selected model or criterion. While these two challenges are the same as those faced in the inference of phylogenetic trees, they are even more daunting
for networks because infinitely many networks are possible for a fixed number of taxa unless constraints on possible network topologies are imposed. In the following sections, we discuss each of these challenges in greater detail.

\subsubsection{Measuring fit to a fixed network}
The first challenge in estimating a phylogenetic network lies in developing methods for determining whether a putative phylogenetic network provides a good fit to the data at hand. In order to make this assessment, it is common to define an optimality criterion that quantifies fit of the network to the data, perhaps under the assumption of a specific evolutionary model. Below, we review the common criteria employed for this purpose, indicating the data types to which they are applied, the evolutionary models assumed, and the software implementations in which they are included.

Before doing so, we note that many of the current methods for network inference assume that data from multiple genes sampled throughout the genome (commonly referred to as {\itshape multilocus data}) are available. In this case, it is necessary to adopt a model for the relationship between the evolutionary histories of the genes and that of the species from which the genes were sampled. The most common model for this purpose is the coalescent model\index{coalescent process} (see, e.g., \cite{kingman1982} for a description of the coalescent process for trees, and \cite{Meng2009,Yu2012,degnan2018} for the extension to phylogenetic networks). In our descriptions below, we note whether multilocus data are assumed and if so, whether the coalescent model is used for inference.

\paragraph{Parsimony.}\index{parsimony} \cite{nakhlehetal2005} proposed a method 
for computing the parsimony score of a phylogenetic network that is formed by adding reticulation events to an existing phylogenetic tree. Their method assumes a single input alignment (and hence a single underlying network), but divides the input data into ``blocks'' in order to evaluate the parsimony score. While \cite{nakhlehetal2005} showed that a polynomial time algorithm for computing the maximum parsimony score can be found when the number of reticulation arcs to be added is fixed, it was subsequently shown that the problem is NP-hard in general \citep{Jin2009}. Note that further extensions of parsimony to phylogenetic networks and the computational complexity of computing the parsimony score have been studied by various authors in recent years (e.g., \cite{Kannan2012,Fischer2015,Bryant2017,VanIersel2017a}).

\paragraph{Minimize deep coalescences.}\index{minimize deep coalescences (MDC)} The mimimize deep coalescences (MDC) method \citep{yuetal2013} was one of the first to be proposed for inference of species-level phylogenetic networks from multilocus data under the coalescent. The method takes as input a collection of gene tree topologies and then defines the score for a putative network as the number of deep coalescent events required to explain the observed gene trees (see \cite{yuetal2013} for details). While the ideas underlying the method are straightforward, it requires that gene trees are first estimated from the alignments for each of the genes. The method is implemented in the PhyloNet software \citep{than2008,wen2018a}. A polynomial-time algorithm for computing the MDC score on level-$1$\index{phylogenetic network!level-$k$!level-$1$} networks has recently been derived \citep{lemayetal2021}.

\paragraph{Likelihood methods.}\index{likelihood} The maximum likelihood framework is the foundation upon which a wide range of statistical methodology is built. Thus, numerous methods for the inference of phylogenetic networks under the maximum likelihood criterion have been proposed. We provide an overview of several current methods categorized by the data type assumed.

\begin{itemize}
    \item {\itshape Single-locus methods.} \cite{Lutteropp2021} recently proposed a method, called NetRAX, to infer networks using maximum likelihood under a single-locus model (i.e., a model that assumes that a single network underlies the entire data set) using standard substitution models.
    \item {\itshape Coalescent methods for multilocus data using gene trees as input.} \cite{yuetal2014} proposed a method for computing the likelihood of a specified phylogenetic network given data consisting of gene trees estimated for a collection of loci under the coalescent model (InferNetwork\_ML in the PhyloNet package). They discuss the computational complexity associated with computation of the likelihood as well as provide methods for assessing confidence. Using the same model as in \cite{yuetal2014}, \cite{kubatko2009} used AIC, AICc, and BIC, measures that are based on the likelihood but employ a ``penalty'' for 
    the addition of parameters to a model, to compare phylogenetic networks with varying numbers of reticulation edges.
\end{itemize}

\paragraph{Pseudolikelihood methods.}\index{likelihood!pseudolikelihood} Due to the complexity of computing the likelihood of a phylogenetic network directly, pseudolikelihood methods have been suggested as an alternative. As in the case of likelihood methods, several methods have been proposed that vary in the type of data they take as input.

\begin{itemize}
\item {\itshape Coalescent methods\index{coalescent process} for multilocus data using gene trees as input.} The PhyloNet package mentioned above includes the InferNetwork\_MPL method \citep{yunakhleh2015} to estimate phylogenetic networks given an input set of gene trees estimated from multilocus data. The pseudolikelihood calculation uses relationships among rooted triples found in the input gene tree topologies. The other popular method in this category is SNaQ, implemented in the PhyloNetworks package \citep{solis-lemus2016,solis-lemus2017}, which computes the pseudolikelihood based on quartet relationships obtained from the input gene trees.

\item {\itshape Coalescent methods for unlinked biallelic markers.} \cite{zhunakhleh2018} proposed a method that uses pseudolikelihood as a criterion to compare phylogenetic networks using sequence data directly when these data consist of a collection of unlinked biallelic markers, such as single nucleotide polymorphisms (SNPs). The method is implemented in the MLE\_BiMarkers module in PhyloNet. This method addresses the fact that using gene trees as input fails to account for variability associated with the initial pre-processing of the data to estimate the gene trees. Because it limits the input data to unlinked biallelic markers, the method is computationally tractable.
\end{itemize}

\paragraph{Bayesian methods.}\index{Bayesian methods} In general, Bayesian methods utilize the likelihood function in computing posterior probabilities of parameters of interest and seek estimators that maximize the posterior probability. In the context of phylogenetic networks, Bayesian methods have an advantage over methods based solely on the likelihood function because in the Bayesian framework (and in particular, through the use of Markov chain Monte Carlo (MCMC) methods to carry out inference; see below), the state space can be augmented to include gene trees for each locus. This allows inference of species-level phylogenetic networks directly from multilocus data without the need to first estimate gene trees for each locus. Methods in this class include SpeciesNetwork in the BEAST2 package \citep{zhangetal2017}, the MSCi model in BPP \citep{flourietal2020}, and MCMC$\_$Seq \citep{wen2018} in PhyloNet.

\subsubsection{Finding optimal phylogenetic networks}
Having described several criteria by which phylogenetic networks can be compared in terms of their fit to empirical data, we now turn our attention to the problem of finding optimal phylogenetic networks under a specified criterion. Below, we discuss several of the associated challenges, including restricting the class of networks considered and development and implementation of methods for efficiently searching the space of networks for those that are optimal.

\paragraph{Restricting network space.} There are two types of restrictions on network space that must be considered when the goal is to find an optimal phylogenetic network. We describe each below.

\begin{itemize}
    \item {\itshape Biologically-motivated restrictions.} Note that in the context of empirical data, reticulation vertices and arcs are assumed to represent evolutionary events, such as hybrid speciation\index{hybridization!hybrid speciation}, gene flow or introgression\index{hybridization!introgression}, LGT\index{lateral gene transfer}, or gene duplication and loss. Thus, the timing and direction of the arcs in a phylogenetic network inferred from empirical data should satisfy the property of time consistency (see p.~\pageref{timeconsistent})\index{phylogenetic network!time-consistent} to insure that horizontal events occur between lineages that exist contemporaneously. Note, however, that the requirement of time consistency depends on the measurement scale. For example, branch lengths measured in coalescent units (number of generations scaled by twice the effective population size) for a network that satisfies the constraint of time consistency will not necessarily lead to a time consistent network in calendar time, because, for example, generation times among lineages might vary substantially. The biological requirement of time consistency leads to a significant scalability challenge in that parameters must be estimated over a constrained parameter space.
  
    \item {\itshape Computationally-motivated restrictions.} For many of the criteria discussed in the previous section (e.g., parsimony and likelihood), addition of horizontal arcs will always improve the value of the objective function. Intuitively, this occurs because each additional arc can explain some portion of the variability in the observed data, thus providing an improvement in the `fit' of the network to the data. For this reason, algorithms for network inference commonly impose limitations on either the number of horizontal arcs or the class of networks over which to search. For example, PhyloNet \citep{yuetal2014} and PhyloNetworks \citep{solis-lemus2016,solis-lemus2017} require the user to specify the number of horizontal arcs (or more precisely, the `maximum number of hybridizations') prior to the search of network space. By running the methods separately with increasing numbers of horizontal arcs, information criteria (i.e., AIC, BIC \citep{kubatko2009}) that penalize for the number of parameters can be used to select among networks with varying numbers of reticulation vertices. PhyloNet \citep{yuetal2014} also implements a cross-validation approach to deal with this problem.

    In addition to the property of time-consistency and the specification of the number of reticulations mentioned above, it is common to place further restrictions on the class of network considered in order the simplify the search. For example, PhyloNetworks \citep{solis-lemus2016,solis-lemus2017} and NANUQ \citep{Allman2019} consider only level-$1$ networks\index{phylogenetic network!level-$k$!level-$1$}.  In contrast, PhyloNet does not provide a clear description of the types of networks considered, while for the SpeciesNetwork module of BEAST2, the birth-hybridization prior assumed limits the class of networks considered to those that can be generated under this process \citep{zhangetal2017}. The extension of the BPP software to handle horizontal events, on the other hand, currently requires that the phylogenetic network be specified in advance, though it can accommodate a relatively broad class of possible networks \citep{flourietal2020}. In reviewing these methods, it was often difficult to determine which classes of networks could be inferred with the various methods. One goal in writing this review is to encourage authors of methods for inferring networks to explicitly state the class of networks on which their methods operate. 
\end{itemize}

\paragraph{Efficiently searching network space.}

\begin{itemize}
    \item {\itshape Criterion-based methods.} Numerous approaches to the problem of finding an optimal network under a selected optimality criterion have been taken, many of which are based on methods that have been applied in the case of searching for optimal phylogenetic trees. For instance, heuristic searches operate by proposing a new network by perturbing a current network using generalizations of move strategies such as nearest neighbor interchange (NNI) or subtree prune and regraft (SPR) to networks, or changing the complexity of the current network by adding or deleting reticulations and then evaluating whether the newly proposed network improves the value of the objective function (for more details on network move strategies, see, e.g., \cite{huber2016b,Gambette2017}; for their implementation in software packages see, e.g., \cite{than2008,solis-lemus2016}). If so, the new network becomes the current network and the process is repeated; if not, a different move is attempted. The process continues until all possible networks have been proposed without any being accepted. Such a strategy clearly provides an uphill search that may result in a network that is only locally optimal. To combat this, heuristic algorithms are often run many times from different initial networks. Another possibility is to apply simulated annealing approaches, which have been successfully used to search for optimal phylogenetic trees (see, e.g., \citep{salter2001,barker2004,stamatakis2005,kubatkoetal2009,stroblbarker2016}).

    \item {\itshape MCMC methods.} Markov chain Monte Carlo algorithms are similar to the searches described above in many ways. For example, they also require a strategy for proposing a new network from an existing network; however, they make a probabilistic decision about acceptance of the proposed network based on the posterior probability. Bayesian methods also provide a more elegant way to handle the problem of inferring the number of horizontal events via reversible-jump MCMC \citep{green1995,green2003,wen2018}. This approach allows the algorithm to \textit{`jump'} dimensions by adding a new reticulation event or by removing an existing reticulation event in a manner that preserves the theoretical property that the chain samples the desired posterior distribution.  While Bayesian algorithms have the appropriate theoretical guarantees, tuning these algorithms to achieve convergence often proves difficult in practice, and such methods are often limited to carrying out inference on only a handful of taxa at a time.

    \item {\itshape Divide-and-conquer methods.} Divide-and-conquer methods, which decompose a problem into smaller sub-problems that can be efficiently solved and whose solutions are then combined to form a solution to the original problem, have also been recently proposed for phylogenetic networks. See \cite{zhu2019} for details.

    \item {\itshape Algorithmic methods.} Algorithmic methods are those that build a phylogenetic network by applying a sequence of deterministic steps. 
For example, methods such as NANUQ \citep{Allman2019} and related methods implemented in the MSCquartets package \citep{rhodes2020} use an algorithmic approach to infer networks under a model that accommodates variation in the evolutionary histories across loci using the coalescent process.
    
\end{itemize}

\subsubsection{Summary}
Inference of phylogenetic networks from genomic data is clearly an important endeavor, but one for which important scalability challenges exist. \cite{hejaseliu2016} have documented the scalability challenges that arise across a range of potential inference methods, and \cite{elworth2019} provide an excellent review of current methods for inferring species-level phylogenetic networks under the coalescent model. This is an active area of research in which we anticipate important conceptual and computational advances in the years to come.  A related challenge is the determination of which network properties can be learned from data, a topic that is discussed in the following section.

\subsection{Identifiability} \label{Sec_Identifiability}
While we have seen in the previous section that the task of estimating phylogenetic networks from data imposes certain scalability challenges, in this section we discuss another challenge in network inference, namely identifiability. Strictly speaking, identifiability refers to a statistical model, where a model parameter is said to be identifiable if any probability distribution arising from the model uniquely determines the value of that parameter. In the setting of phylogenetic network inference, it is thus particularly important that the network parameter (i.e., the network topology and possibly its branch lengths and inheritance probabilities) is identifiable. However, another aspect of identifiability (or rather distinguishability) frequently discussed in relation to phylogenetic network inference is the question of whether a given phylogenetic network is uniquely characterized by certain substructures of the network (e.g., by the phylogenetic trees it `displays') or by certain other structural properties of the network like the distribution of paths in the network or pairwise distances between taxa.

In the following, we discuss both aspects of identifiability (where we distinguish between statistical identifiability and combinatorial identifiability) and relate this to the different phylogenetic network classes introduced earlier.

\subsubsection{Statistical identifiability} \label{subsec_statistical_ident}
An important question in model-based network estimation is the identifiability of the network parameter, i.e., the question of whether the network topology (and possibly additional properties such as branch lengths or inheritance probabilities) can be uniquely identified from data generated by the network. Here, `data generated by the network' typically refers to genomic sequence data observed at the leaves of the network, where sequence evolution along the branches of the network (or along the branches of a set of gene trees associated with the network) is usually modelled as a Markov process\index{Markov model} (for an introduction to Markov processes and their application to phylogenetics, see, for instance, \cite{Steel2016}). 

While several identifiability results for Markov models and the coalescent model on phylogenetic trees have been established in the literature (e.g., \cite{Chang1996,Allman2006,Allman2008,Allman2010,Rhodes2011,Chifman2015,Long2018}), analogous results for phylogenetic networks are much harder to attain. In the following, we summarize some important results obtained in the literature so far. However, we begin by briefly reviewing the notions of identifiability and generic identifiability.

Recall that a model parameter is \emph{identifiable}\index{identifiable} if any probability distribution arising from the model uniquely determines the value of the parameter. As the notion of identifiability is sometimes too strong for practical purposes, generic identifiability\index{identifiable!generic} is often considered instead. Here, a model parameter is said to be \emph{generically identifiable} if the set of parameters from which the original parameter cannot be recovered is a set of Lebesgue measure zero in the parameter space, or in other words, if the parameter is identifiable almost surely.

\paragraph{Identifiability of semi-directed phylogenetic networks.}
While it is natural to model sequence evolution on rooted phylogenetic networks (where sequences evolve from the root of the network towards its leaves), most identifiability results obtained so far focus on \emph{semi-directed phylogenetic networks}, where a semi-directed phylogenetic network\index{phylogenetic network!semi-directed} is obtained from a (directed) rooted phylogenetic network by suppressing the root vertex and undirecting all tree arcs while keeping the direction of all reticulation arcs. This is simply due to the fact that under common Markov models\index{Markov model} of sequence evolution the placement of the root is not identifiable due to the time-reversibility of these models. 

The identifiability of semi-directed phylogenetic networks has been studied by several authors and under different settings. There are both identifiability results assuming the coalescent model as well as results that do not incorporate a coalescent process\index{coalescent process}. 

\begin{itemize}
\item {\itshape Identifiability of topological properties of semi-directed level-$1$ networks assuming a coalescent process.}
\cite{solis-lemus2016} and \cite{Solis-Lemus2020} studied the detectability of hybridization cycles\index{hybridization cycle} and the identifiability of numerical parameters such as branch lengths and inheritance probabilities for semi-directed level-$1$ networks\index{phylogenetic network!level-$k$!level-$1$}\index{phylogenetic network!semi-directed} under a pseudolikelihood model based on the coalescent process. Note that while \cite{solis-lemus2016} provided these results, the mathematical proofs were given later by \cite{Solis-Lemus2020}, who showed that the detectability of hybridization cycles depends on the length of the cycle: cycles containing four or more vertices can be detected from concordance factors (CFs; \cite{Baum2007}), i.e., probabilities of the different quartet topologies displayed on gene trees,  under a pseudolikelihood model, cycles containing two vertices are not detectable, and cycles containing three vertices can be detected under certain conditions. 

In addition and prior to \cite{Solis-Lemus2020}, \cite{Banos2018} provided a mathematical justification for the approach taken in \cite{solis-lemus2016} by showing that most topological properties (in particular, each hybridization cycle of length at least four) of a semi-directed level-$1$ network are generically identifiable from CFs under the network multi-species coalescent model. \index{coalescent process}.

\item{\itshape Identifiability of topological properties of ultrametric level-$1$ networks from log-det distances (assuming a coalescent process).}
Building upon work of \cite{Banos2018}, \cite{Allman2021} showed that most topological properties of ultrametric\index{ultrametric}\footnote{A rooted edge-weighted phylogenetic network $\cN$ is called \emph{ultrametric} if every directed path from the root of $\cN$ to any leaf has the same length.} level-$1$ networks\index{phylogenetic network!level-$k$!level-$1$} can be identified from log-det inter-taxon distances computed from aligned genomic-scale sequences using a combination of the network multispecies coalescent model\index{coalescent process} and a mixture of general time-reversible (GTR) Markov processes\index{Markov model} on gene trees (for details, see \cite{Allman2021}). 

\item {\itshape Identifiability of topological properties of semi-directed networks assuming network-based Markov models.}
While all of the work cited in the previous paragraphs takes into account the coalescent process, other studies have used algebraic approaches to show that certain semi-directed phylogenetic networks can be identified under specific Markov models of sequence evolution. In these studies, it is assumed that all sequence sites have evolved on one of the trees `displayed' (for a formal definition of the concept of a displayed tree\index{displayed tree}, see Section \ref{subsec_combinatorial_ident}) by a network, i.e., Markov models\index{Markov model} on phylogenetic trees are extended to phylogenetic networks by considering a convex combination of the corresponding Markov models on the set of trees displayed by the networks (see, e.g., \cite{Gross2018} for a formal definition of network-based Markov models). For network-based Markov models, the following results were obtained in the literature:
\begin{itemize}
    \item The network parameter (in this case, the network topology) of a network-based Markov model under the Jukes-Cantor \citep{Gross2018}, Kimura 2-parameter, or Kimura 3-parameter \citep{Hollering2021} constraints is generically identifiable with respect to the class of models where the network parameter is a semi-directed network on $n$ leaves with exactly one undirected cycle of length at least four. 
  
    \item The network parameter of a network-based Markov model under the Jukes-Cantor, Kimura 2-parameter, or Kimura 3-parameter constraints is generically identifiable with respect to the class of models where the network parameter is a triangle-free (i.e., each undirected cycle of the network has length at least 4), level-$1$ semi-directed\index{phylogenetic network!level-$k$!level-$1$}\index{phylogenetic network!semi-directed} network on $n$ leaves with $r \geq  0$ reticulation vertices \citep{Gross2020}.
    
    \item Certain semi-directed level-$2$\index{phylogenetic network!level-$k$!level-$2$} networks are identifiable under the Jukes-Cantor, Kimura 2-parameter, and Kimura 3-parameter constraints (for details see~\cite{Ardiyansyah2021}). 
\end{itemize}
\end{itemize}

\paragraph{Identifiability of tree-child networks assuming a probabilistic recombination-mutation model.} Finally, we note
that other extensions to tree-based Markov models are possible. For instance, adapting a model used in \emph{pedigree} reconstruction \citep{Thatte2012}, \cite{Francis2018} introduced an alternative probabilistic recombination-mutation model and established identifiability for almost the entire class of tree-child networks\index{phylogenetic network!tree-child} under this model (and under the mild assumption that the root of the network is not the parent of a reticulation vertex). We refer the reader to \cite{Francis2018} for further details on the model assumptions.

\subsubsection{Combinatorial identifiability} \label{subsec_combinatorial_ident}
In addition to the question of whether a phylogenetic network is identifiable under a certain evolutionary model, it is also of interest to analyze the question of whether a network is identifiable from certain substructures or other structural properties such as inter-taxon distances. 

\paragraph{Encoding networks by subtrees and subnetworks.}
A well-known result in phylogenetics is that a rooted binary phylogenetic tree is uniquely encoded by its set of rooted triples (i.e., the set of induced 3-leaf rooted subtrees) (e.g., \cite{Aho1981,Semple2003}), and so a natural question to ask is whether a rooted binary phylogenetic network can also be uniquely characterized by certain substructures such as subtrees or subnetworks. It turns out that positive results in this regard can be obtained for some of the network classes introduced earlier, whereas arbitrary rooted binary phylogenetic networks are in general not uniquely characterized by simpler substructures. However, before we can elaborate on this, we need to formally define the notion of displayed trees and (displayed) subnetworks.

\begin{figure}[htbp]
    \centering
    \includegraphics[scale=0.2]{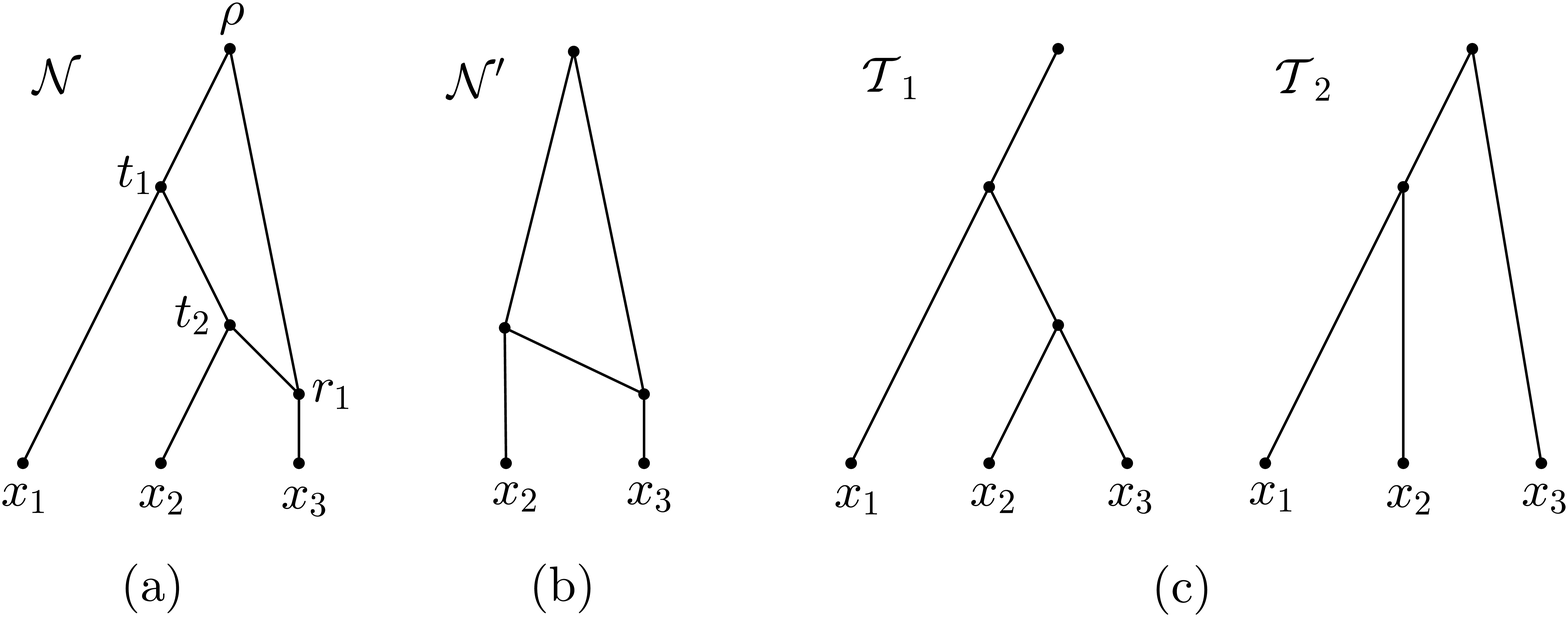}
    \caption{(a) Rooted binary phylogenetic network $\cN$ on $X=\{x_1,x_2,x_3\}$. (b) A subnetwork $\cN'$ on $Y=\{x_2,x_3\}$ displayed by $\cN$. (c) The set $\mathsf{T}(\cN)=\{\cT_1,\cT_2\}$ of phylogenetic $X$-trees displayed by $\cN$.}
    \label{Fig_Display}
\end{figure}

\subparagraph{Displayed trees.\index{displayed tree}}
Let $\cN$ be a phylogenetic network on $X$, and let $\cT$ be a phylogenetic tree on $Y \subseteq X$ (with $Y \neq \emptyset$). Then we say that $\cT$ is displayed\index{displayed tree} by $\cN$ if $\cT$ can be obtained (up to isomorphism) from $\cN$ by deleting arcs and non-root vertices, and suppressing any resulting in-degree one and out-degree one vertices. Note that the roots of $\cN$ and $\cT$ coincide and thus $\rho$ might have out-degree one in $\cT$.\footnote{We remark that sometimes the notion of display is defined as follows: A phylogenetic network $\cN$ displays a phylogenetic tree $\cT$ if $\cT$ can be obtained from $\cN$ be deleting arcs and vertices, and suppressing any resulting in-degree one and out-degree one vertices. In this case, the roots of $\cN$ and $\cT$ are not required to coincide. Note, however, that the two definitions can be used interchangeably. If a tree $\cT$ is displayed by $\cN$ in the first sense, it is also displayed by $\cN$ in the second sense, and vice versa.} 
The set of all phylogenetic $X$-trees displayed by a phylogenetic network $\cN$ on $X$ is denoted as $\mathsf{T}(\cN)$. As an example, the network $\cN$ on $X=\{x_1,x_2,x_3\}$ depicted in Figure~\ref{Fig_Display}(a) displays the two phylogenetic $X$-trees $\cT_1$ and $\cT_2$ depicted in panel (c) of this figure.

\subparagraph{Subnetworks.\index{phylogenetic network!subnetwork}}
Let $\cN$ be a phylogenetic network on $X$ and let $Y \subseteq X$. 
Following the notation of \cite{vanIersel2017}, the \emph{subnet}\index{phylogenetic network!subnetwork} of $\cN$ on $Y$, denoted by $\cN_{\vert Y}$, is defined as the subgraph obtained from $\cN$ by deleting all vertices that are not on any path from the lowest stable ancestor\index{ancestor!lowest stable} $\text{LSA}_{\cN}(Y)$ of $Y$ in $\cN$ to elements in $Y$ and subsequently suppressing all in-degree one and out-degree one vertices and parallel arcs until no such vertices or arcs exist. 
Now, a network $\cN'$ is said to be \emph{displayed}\index{displayed network} by a network $\cN$ if $\cN' = \cN_{\vert Y}$ for some $Y \subseteq X$. As an example, the phylogenetic network $\cN'$ on $Y=\{x_2,x_3\}$ depicted in Figure~\ref{Fig_Display}(b) is displayed by the phylogenetic network $\cN$ on $X=\{x_1,x_2,x_3\}$ depicted in Figure~\ref{Fig_Display}(a). 
Note that, by definition, $\cN_{\vert X} = \cN$ if and only if $\text{LSA}_{\cN}(X) = \rho$. In this case, \cite{vanIersel2017} call $\cN$ a \emph{recoverable} network\index{phylogenetic network!recoverable}.

We are now in a position to summarize some important results on encoding phylogenetic networks by displayed trees and subnetworks.

\begin{itemize}
\item {\itshape Encodings via displayed trees -- arbitrary networks.}
It was shown by \cite{Pardi2015} that arbitrary rooted binary phylogenetic networks on $n$ leaves are not encoded by the set of phylogenetic trees on $n$ leaves they display, even if branch lengths are taken into account \citep[Fig.~3]{Pardi2015}. In particular, there exist non-isomorphic phylogenetic networks $\cN_1$ and $\cN_2$ on $X$ such that $\mathsf{T}(\cN_1) = \mathsf{T}(\cN_2)$, i.e., $\cN_1$ and $\cN_2$ display the same set of phylogenetic $X$-trees. Interestingly, the two networks $\cN_1$ and $\cN_2$ used by \citet[Fig.~3]{Pardi2015} can be distinguished under the network multi-species coalescent model when multiple alleles are sampled per species \citep{Zhu2016}. This nicely illustrates that statistical and combinatorial considerations might not always lead to the same conclusions, and more work on combining the two approaches is needed.

\item {\itshape The NELP property.}\index{NELP property}
\cite{Pardi2015} showed that every phylogenetic network $\cN$ can be transformed into a `canonical' form that displays the same set of phylogenetic trees as $\cN$ and under mild conditions on the branch lengths of $\cN$ (the \emph{no equally long paths} (NELP) property\index{NELP property} which states that no two directed paths in $\cN$ with the same endpoints have the same length, where the length of a path is the sum of the branch lengths assigned to its arcs) this canonical form is unique (up to isomorphism) among all networks satisfying the NELP property. In particular, two networks satisfying the NELP property have the same unique canonical form (up to isomorphism) if and only if they display the same set of trees\index{displayed tree} (with their induced branch lengths). For details, see \cite{Pardi2015}. 

Note that a different canonical form for rooted phylogenetic networks was recently introduced by \cite{Francis2021}. Specifically, \cite{Francis2021} introduced the \emph{normalization} of a phylogenetic network that associates a unique normal network \citep{Willson2009} $\widetilde{\cN}$ \index{phylogenetic network!normal} with any given phylogenetic network $\cN$. 

\item {\itshape Encodings via displayed (caterpillar) trees -- normal networks.}
It was shown by \cite{Willson2011} that normal phylogenetic networks\index{displayed tree}\index{phylogenetic network!normal} on $n$ leaves are uniquely encoded by the set of phylogenetic trees on $n$ leaves they display\label{normalnetworkscaterpillars}. 
More recently, \cite{Linz2020} showed that normal phylogenetic networks are in fact uniquely characterized by their sets of displayed caterpillar trees (particular subtrees that contain precisely one cherry) on three and four leaves. Moreover, \cite{Linz2020} presented a polynomial-time algorithm that takes the set of caterpillar trees on three and four leaves displayed by a rooted binary normal network and reconstructs this network (up to isomorphism). Note that considering caterpillar trees on three and four leaves is essential as there exist two non-isomorphic normal networks that display the same set of rooted triples, i.e., caterpillar trees on three leaves (for an example, see \cite[Fig.~1]{Linz2020}).

\item {\itshape Encodings via rooted triples.}
Phylogenetic networks that are encoded by their rooted triples only seem to be very limited. To our knowledge, the only positive result in this regard was obtained by \cite{Gambette2011} who showed that level-$1$ networks\index{phylogenetic network!level-$k$!level-$1$} are encoded by their sets of displayed triples provided that each reticulation cycle in the network has length at least five.

We remark, however, that even though phylogenetic networks are in general not uniquely encoded by their rooted triples, several algorithms that reconstruct \emph{a} phylogenetic network consistent with a set of triples (consistent in the sense that the triples are displayed by the resulting network) have been developed. We refer the reader to \cite{Poormohammadi2020} for an overview and comparison of different approaches and recent developments.
 
\item {\itshape Encodings via binets, trinets, quarnets, and larger subnetworks.}
Given the limited number of positive results in encoding phylogenetic networks by tree substructures, several studies have analyzed the question whether a phylogenetic network is uniquely characterized by certain network substructures like binets~\citep{Huber2015,vanIersel2017}\index{phylogenetic network!subnetwork!binet}, trinets~\citep{Huber2012, Huber2015,Semple2020, vanIersel2013, vanIersel2017}\index{phylogenetic network!subnetwork!trinet}, and recently also quarnets~\citep{Nipius2020}\index{phylogenetic network!subnetwork!quarnet}, i.e., subnetworks on two, three, and four leaves, respectively. Here, the following positive results have been established:
\begin{itemize}
    \item Recoverable\index{phylogenetic network!recoverable} binary level-$2$ networks\index{phylogenetic network!level-$k$!level-$2$} and binary tree-child networks\index{phylogenetic network!tree-child} are encoded by their sets of displayed trinets\index{phylogenetic network!subnetwork!trinet} \citep{vanIersel2013}. 
    \item Orchard networks \index{phylogenetic network!orchard} are uniquely encoded by their trinets\index{phylogenetic network!subnetwork!trinet} and can be reconstructed in polynomial time from them \citep{Semple2020}. 
    \item Every recoverable\index{phylogenetic network!recoverable} binary level-$3$\index{phylogenetic network!level-$k$!level-$3$} network is encoded by its set of displayed quarnets\index{phylogenetic network!subnetwork!quarnet} \citep{Nipius2020}.
\end{itemize}

\noindent However, as far as arbitrary phylogenetic networks are concerned, it has been shown by \cite{Huber2014} that even if \emph{all} subnetworks induced on all proper subsets of the leaves of some rooted binary phylogenetic network are given, the network is still not necessarily determined by this information.  \vspace{0.05in}

\item {\itshape Encodings via reticulate-edge-deleted subnetworks.}
\cite{Murakami2019} recently showed that level-$k$ tree-child networks\index{phylogenetic network!tree-child}\index{phylogenetic network!level-$k$} with $k \geq 2$ can be determined and reconstructed in polynomial time from their \emph{reticulate-edge-deleted subnetworks}\index{phylogenetic network!subnetwork!reticulate-edge-deleted}, which are subnetworks obtained by deleting a single reticulation arc. Even stronger, level-$k$ tree-child networks with $k \geq 2$ are encoded by their subnetworks obtained from deleting one reticulation arc from each biconnected component with $k$ reticulations (for details see \cite{Murakami2019}).

\item {\itshape Encodings via tri-LGT-nets.}
\cite{Cardona2017} showed that a subclass of time-consistent LGT networks\index{phylogenetic network!LGT network} (referred to as time-consistent \emph{BAN-LGT networks}) can be uniquely reconstructed (up to redundant arcs (shortcuts) and isomorphism) from the set of their \emph{tri-LGT-nets}, i.e., from the set of their induced 3-leaf subnetworks  (for further details see \cite{Cardona2017}).
\end{itemize}

\paragraph{Encoding networks by sets of paths and pairwise distances.} 
An alternative approach to encoding a phylogenetic network by substructures is to consider other structural properties of the network such as the distribution of path lengths or pairwise distances between the leaves. 
In the following, we review different concepts and ideas used in this regard.

\begin{itemize}
\item {\itshape The $\mu$-representation of phylogenetic networks.}
The $\mu$-representation of phylogenetic networks relies on the notion of path-multiplicity vectors introduced by \cite{Cardona2009}. Using a similar notation as \cite{Cardona2009}, let $\cN=(V,E)$ be a rooted phylogenetic network on $X= \{x_1, \ldots, x_n\}$. For every vertex $v \in V$ and $i \in \{1, \ldots, n\}$, let $m_i(v)$ denote the number of different paths from $v$ to leaf $x_i$. Then, the \emph{path-multiplicity vector}, or \emph{$\mu$-vector}\index{phylogenetic network!$\mu$-vector} for short, of $v \in V$ is defined as $\mu(v) = \left(m_1(v), \ldots, m_n(v) \right)$,
i.e., $\mu(v)$ is an $n$-tuple containing the number of paths from $v$ to each leaf of $\cN$. Now, the \emph{$\mu$-representation}\index{phylogenetic network!$\mu$-representation} of a phylogenetic network $\cN=(V,E)$ is the multiset $\mu(\cN)$ of $\mu$-vectors of its vertices. More precisely, the elements of this multiset are the vectors $\mu(v)$ with $v \in V$, and the multiplicity of each element is the number of vertices having this element as its $\mu$-vector.

Based on this, \cite{Cardona2009} showed that two (not necessarily binary) tree-child networks\index{phylogenetic network!tree-child} $\cN_1$ and $\cN_2$ are isomorphic if and only if they have the same $\mu$-representation, i.e., if and only if $\mu(\cN_1) = \mu(\cN_2)$. Moreover, given the $\mu$-representation of a phylogenetic network $\cN$, the network can be reconstructed in polynomial time. For further details, see \cite{Cardona2009}.

In addition, \cite{Cardona2008} showed that the same is true for semi-binary\index{phylogenetic network!semi-binary}\footnote{In a semi-binary phylogenetic network all reticulation vertices have in-degree precisely two, but tree vertices may have an out-degree strictly greater than two. Note that every binary phylogenetic network is in particular a semi-binary phylogenetic network.} time-consistent tree-sibling\index{phylogenetic network!tree-sibling time-consistent} phylogenetic networks. 

Recently, these results were extended to the larger class of orchard phylogenetic networks by \cite{Erdos2019} and \cite{Bai2021}\index{phylogenetic network!orchard}. More precisely, \cite{Bai2021} showed that any (not necessarily binary) stack-free\index{phylogenetic network!stack-free} orchard phylogenetic network is uniquely encoded (up to isomorphism) by its $\mu$-representation\index{phylogenetic network!$\mu$-representation} (called `ancestral profile' therein). 

Note that while the results by \cite{Cardona2008,Cardona2009} entail uniqueness within the class of time-consistent tree-sibling, respectively tree-child, networks, the result of \cite{Bai2021} proves uniqueness among all rooted phylogenetic networks.

In addition, \cite{Bai2021} showed that if the `stack-free' condition is omitted, the ancestral profile of an orchard network $\cN$ uniquely encodes $\cN$ within the class of orchard networks up to the resolution of vertices of high in-degree. For further details, see \cite{Bai2021}. 

\item {\itshape Encoding phylogenetic networks by pairwise distances.}
Some phylogenetic networks can be encoded by considering pairwise distances between the leaves or taxa of the network. 
Here, distances can either be measured in terms of topological path lengths between leaves, where the length of a path is defined as the number of arcs contained in it, or in terms of the sum of edge weights on these paths when the network is equipped with branch lengths. An important concept in both cases is the notion of \emph{up-down paths}\index{up-down-path} introduced by \cite{Bordewich2015}.

\begin{description}
\item[Up-down paths.]
Let $\cN$ be a rooted phylogenetic network on $X$. Then, following the notation of \cite{Bordewich2015}, an \emph{underlying path} of $\cN$ is a path of the undirected graph containing undirected edges of arcs of $\cN$. Now, for any two elements $x,y \in X$, an \emph{up-down path} from $x$ to $y$ is an underlying path $(x, v_1, v_2, \ldots, v_{k-1}, y)$ in $\cN$ such that, for some $i \leq k-1$, the network $\cN$ contains the arcs
\[ (v_i,v_{i-1}), (v_{i-1},v_{i-2}), \ldots, (v_1,x)\]
and 
\[ (v_i,v_{i+1}), (v_{i+1}, v_{i+2}), \ldots, (v_{k-1},y).\] 
As an example, in Figure~\ref{Fig_Display}(a), $(x_1, t_1, t_2, r_1, x_3)$ is an up-down path in $\cN$ from $x_1$ to $x_3$.

\item[Unweighted phylogenetic networks.]
Based on the notion of up-down-paths, \cite{Bordewich2015} showed that unweighted binary tree-child\index{phylogenetic network!tree-child} networks with no arc between the two children of the root can be reconstructed (up to isomorphism) from the \emph{multi-set} of distances between taxa, where the distance between two taxa, say $x$ and $y$, is measured in terms of the number of arcs on the up-down-paths from $x$ to $y$, and so can all binary time-consistent networks\index{phylogenetic network!time-consistent} with no `crowns'\footnote{Let $\cN$ be a phylogenetic network. Then, a \emph{crown} is an (undirected) cycle in $\cN$ consisting only of reticulation arcs.}, no arc between the two children of the root, and all reticulation vertices being visible. 

Moreover, \cite{Bordewich2015} showed that binary time-consistent tree-child\index{phylogenetic network!tree-child time-consistent} networks can be reconstructed (up to isomorphism) from the \emph{set} of distances between taxa in polynomial time. 

\item[Edge-weighted rooted phylogenetic networks.]
In the case of edge-weighted phylogenetic networks, various results have been obtained in the literature.

First, improving earlier results on the reconstructability of ultrametric\index{ultrametric} galled networks\index{phylogenetic network!galled network} \citep{Chan2005} and networks with a single reticulation cycle \citep{Willson2013}\footnote{Note that \cite{Willson2013} considered the problem of reconstructing a phylogenetic network $\cN$ given the \emph{tree-average distance}~\citep{Willson2012} between any pair of taxa, which is the expected value of their distance in the trees displayed by $\cN$, where each displayed tree has a certain probability obtained from assigning inheritance probabilities\index{inheritance probability} to the reticulation arcs of $\cN$. \label{footnote_avgdist}}, \cite{Bordewich2016a} introduced a polynomial-time algorithm that reconstructs an ultrametric tree-child\index{phylogenetic network!tree-child} network from the set of distances between each pair of taxa, where the set of distances between a pair of taxa, say $x$ and $y$, is the set of the lengths of the up-down-paths\index{up-down-path} from $x$ to $y$ (where the length of any such a path is the sum of branch lengths of the edges in this path rather than the number of edges in the path). 

\cite{Bordewich2016a} introduced the algorithm \textsc{NetworkUPGMA} that takes a 2-dimensional array of sets of distances (where distances between taxa may for example reflect evolutionary distance estimated from genetic sequence data) and returns an ultrametric tree-child network displaying the distance data if such a network exists.

In a subsequent paper, \cite{Bordewich2017} showed that any tree-child network $\cN$ on $X$ with an outgroup (i.e., an element $x \in X$ adjacent to the root of $\cN$) and strictly positive branch lengths is essentially encoded by the multi-set of distances between all pairs of taxa (again, distance refers to sum of branch lengths in up-down-paths) provided that for each reticulation vertex $r$, both reticulation arcs directed into $r$ are of equal length. Note that `essentially encoded' refers to the fact that this encoding is unique up to re-weighting the edges at the root and at each reticulation (for technical details and examples see \cite{Bordewich2017}). Moreover, \cite{Bordewich2017} introduced a polynomial-time algorithm for reconstructing edge-weighted tree-child networks from inter-taxa distance data. Note, however, that in a tree-child network, the size of the collection of inter-taxon distances can be exponential in the number of leaves of the network. 

In a recent paper, \cite{Bordewich2018a} showed that for normal phylogenetic networks the same results are obtained with only a quadratic number of inter-taxon distances by using the shortest distance between any pair of taxa (i.e., the sum of branch lengths in a shortest up-down path between the two taxa).

\item[Edge-weighted semi-directed networks.] Even more recently, \cite{Xu2021} returned to the identifiability of local and global properties of edge-weighted phylogenetic networks from \emph{average pairwise distances} (\cite{Willson2012}; see also Footnote \ref{footnote_avgdist}). Importantly, the authors show that root location and lengths of reticulation arcs are generally \emph{not} identifiable from average distances, and then focus on the identifiability of ``zipped-up semi-directed'' networks, where a network is zipped-up if all its reticulation arcs have length zero. For networks of this type, several positive and negative results regarding their identifiability from average distances are obtained and additional conjectures are posed. We refer the reader to \cite{Xu2021} for further details.
\end{description}
\end{itemize}

\subsubsection{Summary}
\noindent In summary, several results concerning the identifiability of phylogenetic networks have been established in recent years and more are likely to be obtained in the near future. However, as the previous paragraphs showed, positive results can mostly only be obtained for restricted network classes, but not for arbitrary phylogenetic networks. While this is to be expected (given the potential complexity of arbitrary phylogenetic networks) it is important to keep these limitations in mind when, for example, devising new network inference methods. \\

\section{Concluding remarks} \label{Sec_Conclusion}
The aim of the present manuscript is to provide a thorough and comprehensive review of the multitude of different classes of rooted binary phylogenetic networks defined in the mathematical literature. We have reviewed and discussed their structural properties and indicated their biological interpretation whenever possible. For some network classes, for instance temporal networks or LGT networks, it is straightforward to provide a biological interpretation, whereas for other network classes, the biological meaning is less evident. This is to be expected, however, as many of the structural constraints that have been considered in the literature were not introduced to model certain biological processes, but rather to simplify  mathematical and computational analyses so that problems related to the inference of phylogenetic networks would become tractable. 

We have noted that imposing structural constraints on  network topologies is often an important step in addressing the scalability and identifiability challenges faced in estimating phylogenetic networks from data, and hope that our review of possible classes of networks will encourage those who develop such methods to carefully describe the class of networks considered by their methods. Even though positive results concerning scalability and identifiability are currently limited to a handful of the more than 20 network classes we have discussed, we expect constant progress in the future as the study and estimation of phylogenetic networks is a growing field of research. In addition, we remark that while we could not find a biological interpretation for all of the network classes discussed, we do not claim that such an interpretation is non-existent. It could well be the case that network types that seem to lack an underlying biological principle in the context of our current understanding of evolution will turn out to be biologically meaningful as this understanding expands.

\section*{Acknowledgements}
JCP was supported by the Ministerio de Ciencia e Innovación (MCI), the Agencia Estatal de Investigación (AEI) and the European Regional Development Funds (ERDF); through project PGC2018-096956-B-C43 (FEDER/MICINN/AEI).
KW  was  supported  by  The  Ohio  State  University’s President’s Postdoctoral Scholars Program. All authors thank two anonymous reviewers for detailed comments on an earlier version of this manuscript.

\bibliography{NetworkClasses.bib}

\printindex
\end{document}